\documentclass[pra,showpacs,twocolumn]{revtex4-1}

\usepackage{psfrag}
\usepackage{bm,graphicx,amsmath}
\usepackage{color}
\newcommand{\vc}[1]{\mathbf{#1}}
\begin{document}

\title{Excitation dynamics in a lattice Bose gas\\ within the time-dependent
Gutzwiller mean-field approach}

\author{Konstantin V.~Krutitsky$^1$ and Patrick Navez$^{1,2}$}

\affiliation
{
$^1$Fakult\"at f\"ur Physik der Universit\"at Duisburg-Essen, Campus Duisburg,
Lotharstra{\ss}e 1, 47048 Duisburg, Germany\\
$^2$Institut f\"ur Theoretische Physik, TU Dresden, 01062 Dresden, Germany
}

\date{\today}

\begin{abstract}
The dynamics of the collective excitations of a lattice Bose gas
at zero temperature is systematically investigated
using the time-dependent Gutzwiller mean-field approach.
The excitation modes are determined within the framework of the linear-response theory
as solutions of the generalized Bogoliubov-de~Gennes equations valid in the
superfluid and Mott-insulator phases at arbitrary values of parameters.
The expression for the sound velocity derived in this approach coincides with
the hydrodynamic relation.
We calculate the transition amplitudes for the excitations in the Bragg scattering process
and show that the higher excitation modes give significant contributions.
We simulate the dynamics of the density perturbations and show that their propagation
velocity in the limit of week perturbation is satisfactorily described by the predictions
of the linear-response analysis.
\end{abstract}

\pacs{03.75.-b,03.75.Lm,05.45.Yv}

\maketitle

\section{Introduction}

Studies of excitations of ultracold atoms in optical lattices play an important role
for understanding of their physical properties and dynamical behavior.
For instance, celebrated quantum phase transition from the superfluid (SF) into
the Mott-insulator (MI)~\cite{Fisher} is
accompanied by the opening of the gap in the excitation spectrum~\cite{Fisher}.
In the deep optical lattices the system of interacting bosons is satisfactorily described
by the Bose-Hubbard model~\cite{Fisher,JBCGZ}.
Since the model is not integrable, exact analytical results can be obtained only
in few special cases like in the limit of weakly interacting gas~\cite{Rey03}
or for hard-core bosons in one dimension~\cite{Cazalilla}.
However, in general exact results can be obtained only with the aid of numerical methods.

Exact numerical results for the spectrum of low-energy excitations were obtained
by means of diagonalizations of the Bose-Hubbard Hamiltonian~\cite{RB2003R,RB2003,RB2004}.
Although numerical diagonalization can be really done only for rather small lattices
which are far from the thermodynamic limit, this allows to capture all the main characteristic features
of realistic systems.
Numerical results for larger systems of the same size as in real experiments
with ultracold atoms~\cite{Greiner2002}
have been obtained by quantum Monte Carlo methods~\cite{KK09}
which allow to compute the spectral properties~\cite{PEH,CSPS08}.
Simulation of the real-time dynamics of quantum systems within quantum Monte Carlo is also
possible (see, e.g.,~\cite{tQMC})
but has not been yet performed for Bose systems.
Ground-state properties as well as the real-time dynamics of one-dimensional
systems subject to external perturbation were studied, e.g.,
in Refs.~\cite{KSDZ04,KSDZ05,PRA09,EFG11}
by the powerful numerical density-matrix renormalization group (DMRG) method
giving an access to the excitation spectrum~\cite{Sch2005}
but so far restricted to one-dimensional systems.


Mean-field theories in the dimensions higher than one allow self-consistent study of
the excitations and dynamics in the lowest order with respect to quantum fluctuations.
In a weakly interacting regime, the atoms are fully condensed
and the system is satisfactorily described
by the time-dependent discrete Gross-Pitaevskii equation (DGPE).
The excitation spectrum can be calculated using Bogoliubov-de~Gennes (BdG)
equations~\cite{stringari,MKP03,Menotti04}.
It has a form of the Goldstone mode characterized at large wavelengths by the sound velocity $c$.
The latter is related to the compressibility $\kappa$, the effective mass $m^*$
and the condensate density $\left|\psi\right|^2$
through the relation~\cite{MKP03}
\begin{equation}
c
=
\sqrt{\left|\psi\right|^2/\kappa m^*}
\;.
\label{vs}
\end{equation}
The same expression for the sound velocity can be also derived
from the Bogoliubov theory in the operator formalism~\cite{Rey03} and
from the hydrodynamic approach~\cite{Fisher,Dalibard}.


In the strongly interacting regime, condensate fraction becomes suppressed and
for commensurate fillings the system can undergo a transition
from the SF into the MI state~\cite{Fisher,Dalibard,LSADSS}.
In this regime, DGPE is not valid and on the mean-field level must be replaced by more general Gutzwiller equations~(GE) which are exact for a gas of infinite dimensions~\cite{Fisher,Rokshar,Krauth,Navez}.
GE were successfully used to study various phenomena like
creation of molecular condensate by dynamically melting a MI~\cite{JVCWZ02},
dynamical transition from the SF to Bose-glass phase due to controlled growing
of the disorder~\cite{DZSZL03},
the gas dynamics in time-dependent lattice potentials~\cite{Z},
transport of cold atoms induced by the shift of the underlying harmonic potential~\cite{SH07},
dynamics of metastable states of dipolar bosons~\cite{TML08},
and the soliton propagation~\cite{KLL10}.


Excitations above the ground state described by the GE were studied
using the random phase approximation (RPA)~\cite{SKPR,Oosten2,Gerbier,Menotti08,HM},
the Schwinger-boson approach~\cite{Huber},
time-dependent variational principle with subsequent quantization~\cite{HTAB},
Hubbard-Stratonovich transformation~\cite{Oosten,SD05},
slave-boson representation of the Bose-Hubbard model~\cite{DODS03},
standard-basis operator method~\cite{KNN06,OKM06},
and Ginzburg-Landau theory~\cite{Pelster}.
All these methods show that
in the MI phase the spectrum of excitations consists of the
particle- and hole-modes, both with nonvanishing gaps~\cite{Huber,KPS}.
In the SF phase, the lowest branch is a gapless Goldstone mode~\cite{Huber,HTAB,KPS,Bissbort},
while higher branches are gapped~\cite{Huber,Bissbort}.
However, different approximations used in the calculations by different methods
do not always lead to the same final results.
For instance, in Ref.~\cite{Menotti08} it was checked numerically that the RPA gives the same result for the sound velocity as Eq.~(\ref{vs}),
whereas the analytical expressions derived in Refs.~\cite{SD05,KPS07} differ from that.

Excitations above the Gutzwiller ground state can be also investigated using
generalization of the BdG equations directly derived from the GE within the framework
of the linear response theory. This method was used
for the lattice Bose gas with short-range~\cite{KPS07,Bissbort}
as well as long-range~\cite{KPS,TML08} interactions.
This approach, which was not so far widely used to study the lattice Bose gas
allows to obtain results consistent with other mean-field approaches mentioned above and to study
the ground state, stationary excitation modes as well as the dynamics of the gas
on equal footing.


The excitations can be probed in experiments on inelastic light scattering (Bragg spectroscopy)
which provide an information on the dynamic structure factor and one-particle spectral function.
Recently such experiments were carried out with ultra-cold rubidium atoms in optical lattices
of different dimensions in the SF phase~\cite{Sengstock1,Bissbort,Heinze}
and across the SF-MI transition~\cite{Inguscio,Inguscio2010}.
Theoretical analysis has been developed using
exact diagonalization of the Bose-Hubbard Hamiltonian~\cite{Clark,Batrouni},
quantum Monte Carlo simulations in one dimension~\cite{PEH,Batrouni},
perturbation theory valid deep in the MI phase~\cite{Clark},
hydrodynamic theory~\cite{Machida},
exact Bethe-ansatz solution of the Lieb-Liniger model~\cite{CC06},
extended fermionization~\cite{Batrouni},
RPA~\cite{Menotti08}.
Analogous studies were also performed within the Gutzwiller approximation~\cite{Oosten2,Bissbort,Huber}.
However, previous calculations are valid only for the MI~\cite{Oosten2,Huber}
or close to it~\cite{Huber}.
It was argued that the second excitation branch in the SF cannot be detected
by the Bragg spectroscopy in the linear regime~\cite{HTAB}.
In Ref.~\cite{Bissbort}, the calculations beyond the linear response theory
taking into account the harmonic trapping potential and the finite duration
of the probing Bragg pulse were performed which are in good quantitative agreement
with the experimental data reporting the observation of the second excitation branch.
In the present work, we will study in details the possibility to observe the second excitation
mode in a homogeneous lattice in the linear-response regime.


Experimentally sound waves can be observed with the aid of an external potential
which creates a density perturbation of the gas.
Corresponding numerical simulations for the lattice gas were performed on the basis
of the DGPE~\cite{Menotti04} and for soft-core bosons in 1D making use of
the DMRG method~\cite{KSDZ05}.
The sound velocity extracted from these simulations is in perfect agreement with
Eq.~(\ref{vs}) in the case of the DGPE~\cite{Menotti04}
and has a correct asymptotic behavior in 1D in the limits of weak
and strong interactions~\cite{KSDZ05}, where analytical expressions are known.


The purpose of this paper is to give a comprehensive self-consistent description of the
collective excitations as well as experimental techniques for their observation
within the time-dependent Gutzwiller ansatz which is gapless and satisfies
the basic conservation laws, in particular, f-sum rule.
Solution of GE allows not only to obtain the excitation dispersion relations
but also to calculate the transition amplitudes in the Bragg scattering process.
We present a derivation of Eq.~(\ref{vs}) from the generalized BdG equations.
Furthermore, the time-dependent approach is also used to investigate
the sound wave propagation in the case of a stronger
perturbation generated by switching off a local potential.
In this way, we can determine the speed at which this perturbation propagates
and compare with the predictions of the linear-response theory.
We emphasize that the Gutzwiller ansatz is the only approximation used in the present work
and the results are valid in the whole range of parameters.

The paper is organized in the following manner.
In Sec.~\ref{GA}, we present the time-dependent GE.
Their ground state solutions are discussed in Sec.~\ref{GS}.
In Sec.~\ref{E}, we determine the spectrum of collective excitations
using Gutzwiller-Bogoliubov-de Gennes equations.
Sec.~\ref{T} is devoted to the Bragg scattering.
In Sec.~\ref{SW}, we simulate the sound-wave propagation.
The conclusions are presented in Sec.~\ref{C}.

\section{\label{GA}The time-dependent Gutzwiller ansatz }

We consider a system of ultracold interacting bosons in a
$d$-dimensional isotropic lattice described by the Bose-Hubbard Hamiltonian
\begin{eqnarray}
\label{BHH}
\hat H
&=&
-J
\sum_{\bf l}
\sum_{\alpha=1}^d
\left(
    \hat a_{\bf l}^\dagger
    \hat a_{{\bf l}+{\bf e}_\alpha}
    +
    {\rm h.c.}
\right)
\nonumber\\
&+&
\frac{U}{2}
\sum_{\bf l}
\hat a^\dagger_{\bf l}
\hat a^\dagger_{\bf l}
\hat a_{\bf l}
\hat a_{\bf l}
-
\mu
\sum_{\bf l}
\hat a^\dagger_{\bf l} \hat a_{\bf l}
\;,
\end{eqnarray}
where the site index ${\bf l}$ is a $d$-dimensional vector,
${\bf e}_\alpha$ is a unit vector on the lattice in the direction $\alpha$,
$J$ is the tunneling matrix element,
$U$ is the repulsive on-site atom-atom interaction energy,
and $\mu$ the chemical potential.
The annihilation and creation operators at site ${\bf l}$,
$\hat a_{\bf l}$ and $\hat a^\dagger_{\bf l}$,
obey the bosonic commutation relations.
The momentum operator
\begin{displaymath}
\hat{\bf P}
=
i P_0
\sum_{\alpha=1}^d
{\bf e}_\alpha
\sum_{\bf l}
\left(
    \hat a_{{\bf l}}^\dagger
    \hat a_{{\bf l}+{\bf e}_\alpha} - {\rm h.c.}
\right)
\;,
\end{displaymath}
where $P_0$ is a constant determined by the parameters of the periodic potential
creating the lattice, does not commute with the Hamiltonian~(\ref{BHH}) due to the interaction term.
Instead,  the quasi-momentum operator defined as
\begin{displaymath}
\hat{\bf p}
=
\sum_{\bf k}
{\bf k}
\hat a_{\bf k}^\dagger
\hat a_{\bf k}
\;,\quad
\hat a_{\bf k}
=
\sum_{\bf l}
e^{-i{\bf k}\cdot{\bf l}}\hat a_{{\bf l}}/L^{d/2}
\;,
\end{displaymath}
where $k_\alpha=2\pi n_\alpha/L$, $n_\alpha=0,\dots,L-1$,
with $L$ being the number of lattice sites in each spatial direction,
commutes with the latter.

Our analysis employs the Gutzwiller ansatz. Thereby, eigenstates of
the Hamiltonian~(\ref{BHH}) are taken as tensor products of local states
\begin{equation}
\label{state}
|\Phi\rangle
=
\bigotimes_{\bf l}
|s_{\bf l}\rangle
\;,\quad
|s_{\bf l}\rangle
=
\sum_{n=0}^\infty
c_{{\bf l},n}
|n\rangle_{\bf l}
\;.
\end{equation}
where $|n\rangle_{\bf l}$ is the Fock state with $n$ atoms at site ${\bf l}$.
Normalization of the $|s_{\bf l}\rangle$ imposes
\begin{equation}
\sum_{n=0}^\infty
\left|
    c_{{\bf l},n}
\right|^2
=
1
\;.
\nonumber
\end{equation}
The mean number of condensed atoms in this model is given by $|\psi_{\bf l}|^2$,
where
\begin{equation}
\label{psi}
\psi_{\bf l}
=
\langle \hat a_{\bf l}\rangle
=
\sum_{n=1}^\infty
c_{{\bf l},n-1}^* c_{{\bf l},n}
\sqrt{n}
\end{equation}
is the condensate order parameter. One can easily show that $|\psi_{\bf l}|^2$
cannot be larger than the mean occupation number
\begin{equation}
\langle\hat n_{\bf l}\rangle
=
\sum_{n=1}^\infty
n
\left|
    c_{{\bf l},n}
\right|^2
\;.
\end{equation}
Minimization of the functional
\begin{displaymath}
i\hbar
\sum_{n=0}^\infty
(c_{{\bf l},n}^*
\partial_t c_{{\bf l}n}
-
c_{{\bf l}n}
\partial_t c^*_{{\bf l}n})
-
\langle H \rangle
\end{displaymath}
leads to the system of GE~\cite{Z,BPVB07}:
\begin{eqnarray}
\label{GEd}
i\hbar
\frac{d c_{{\bf l}n}}{dt}
&=&
\sum_{n'}
H_{\bf l}^{n n'}
c_{{\bf l}n'}
\;,
\\
H_{\bf l}^{n n'}
&=&
\left[
    \frac{U}{2}n(n-1)
    -
    \mu n
\right]
\delta_{n',n}
\nonumber\\
&-&
J
\sqrt{n'}
\delta_{n',n+1}
\sum_{\alpha=1}^d
\left(
    \psi^*_{{\bf l}+{\bf e}_\alpha}
    +
    \psi^*_{{\bf l}-{\bf e}_\alpha}
\right)
\nonumber \\
&-&
J
\sqrt{n}
\delta_{n,n'+1}
\sum_{\alpha=1}^d
\left(
    \psi_{{\bf l}+{\bf e}_\alpha}
    +
    \psi_{{\bf l}-{\bf e}_\alpha}
\right)
\;.
\nonumber
\end{eqnarray}
The Gutzwiller approximation is {\it conserving} since
these equations do not violate conservation laws of the original Bose-Hubbard model.
The expectation values of the quasi-momentum, total energy, and the total number of
particles remain constant in time.

As it follows from the form of the state (\ref{state}), Gutzwiller approximation
neglects quantum correlations between different lattice sites but takes into account
on-site quantum fluctuations. This appears to be enough for satisfactory description
of the SF-MI quantum phase transition.
Due to the fact that in the equations of motion (\ref{GEd})
the coefficients $c_{{\bf l}n}$ for different sites are coupled to each other,
Gutzwiller ansatz can be also used to study the dynamics of excitations.

>From (\ref{GEd}), we deduce the following equation for the order parameter:
\begin{eqnarray}
\label{GEd2}
i\hbar
\frac{d \psi_{\bf l}}{dt}
=
&-&
J
\sum_{\alpha=1}^d
\left(
    \psi_{{\bf l}+{\bf e}_\alpha}
    +
    \psi_{{\bf l}-{\bf e}_\alpha}
\right)
-\mu \psi_{\bf l}
\nonumber \\
&+&
U
\sum_{n=0}^\infty
(n-1)\sqrt{n}
c_{{\bf l},n-1}^*  c_{{\bf l},n}
\;.
\end{eqnarray}
This equation becomes closed if we assume the coherent state
\begin{equation}
\label{cs}
c_{{\bf l},n}
=
c^{coh}_{{\bf l},n}
\equiv
\exp
\left(
    -|\psi_{\bf l}|^2/2
\right)
\psi_{\bf l}^n/\sqrt{n!}
\;,
\end{equation}
which is an exact solution of Eq.~(\ref{GEd}) for $U=0$.
In this case, the replacement of the last
term of Eq.~(\ref{GEd2}) by this distribution leads to the result
\begin{equation}
U\sum_{n=0}^\infty
(n-1)\sqrt{n}
({c^{coh}_{{\bf l},n-1}})^*
c^{coh}_{{\bf l},n}
=
U|\psi_{\bf l}|^2 \psi_{\bf l}
\;,
\nonumber
\end{equation}
so that we recover the DGPE valid for small $U/J$.

\section{\label{GS}Ground state}

\begin{figure}[t]

\hspace{-2.5cm}
\includegraphics[width=10cm]{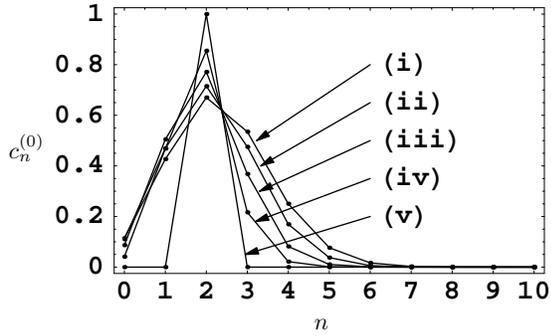}

\caption{Ground-state solutions for the atomic distribution $c_n^{(0)}$.
The scaled chemical potential $\mu/U=1.2$
and the tunneling rates $2dJ/U$:
0.7~(i), 0.5~(ii), 0.3~(iii), 0.15~(iv), and 0.05~(v).
The lines connecting the dots are to guide the eye.}
\label{cn}
\end{figure}

\begin{figure}[t]



\hspace{-3cm}
\includegraphics[width=10cm]{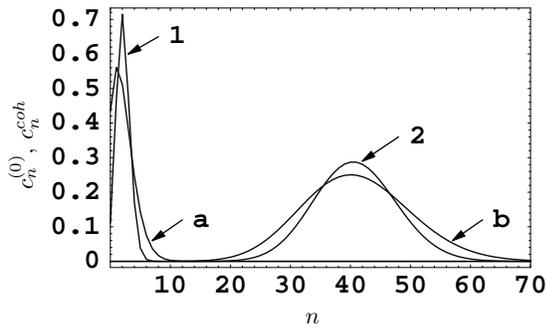}

\caption{Comparison of $c_n^{(0)}$~(1,2) with $c_n^{coh}$~(a,b).
The scaled chemical potential $\mu/U=1.2$
and the tunneling rates $2dJ/U$: 0.5~(1,a), 50~(2,b).
}
\label{cn_cs}
\end{figure}

In the ground state, the coefficients $c_{{\bf l},n}$
do not depend on the site index ${\bf l}$ so that
the solution has the form
\begin{equation}
c_{{\bf l},n}(t)
\equiv
c_n^{(0)}
\exp
\left(
    -i \omega_0 t
\right)
\;,
\end{equation}
so that
$
\langle
     \hat n_{\bf l}
\rangle
\equiv
\langle
     \hat n
\rangle
$, where $\hat n= \sum_\vc{l}\hat n_{\bf l}/L^d$.
The coefficients $c_n^{(0)}$ are calculated numerically
solving Eq.~(\ref{GEd}) by means of exact diagonalization in the same manner
as in Refs.~\cite{SKPR,Oosten}.
The results are shown in Figs.~\ref{cn} and \ref{cn_cs}.
$c_n^{(0)}$ has a broad distribution in the SF phase, where the corresponding
$\psi^{(0)}=\psi_{\bf l}$ defined by Eq.~(\ref{psi}) does not vanish.
In the MI phase, however,
\begin{equation}
\label{gsmi}
c_n^{(0)}=\delta_{n,n_0}
\end{equation}
resulting in $\psi^{(0)}=0$.
$\omega_0$ is determined in both phases from:
\begin{equation}
\hbar\omega_0
=
- 4dJ
\psi^{(0)2}
+
\sum_{n=0}^\infty
\left[
    \frac{U}{2}n(n-1) - \mu n
\right]
c_n^{(0)2}
\;.
\end{equation}

\begin{figure}[t]

\hspace{-2.5cm}
\includegraphics[width=10cm]{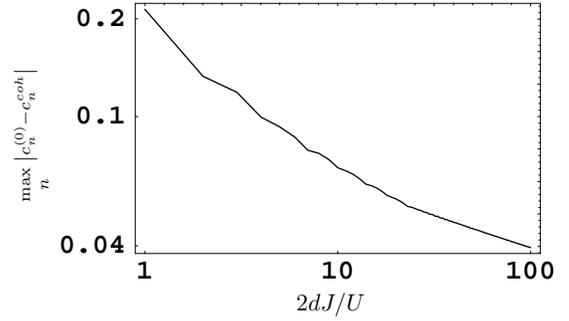}

\caption{
Deviation of the exact ground state from the coherent state for
the scaled chemical potential $\mu/U=1.2$.
}
\label{deviation}
\end{figure}

In Figs.~\ref{cn_cs} and~\ref{deviation},
we compare the coefficients $c_n^{(0)}$ obtained by numerical
solution of Eqs.~(\ref{GEd}) with the corresponding results for the coherent state
(\ref{cs}). The value of $\psi^{(0)}$ for Eq.~(\ref{cs}) was calculated according to
Eq.~(\ref{psi}) using $c_n^{(0)}$.
These coefficients  converge towards a coherent state distribution for increasing $J/U$
according to a power law (Fig.~\ref{deviation}), thus, justifying the use
of the DGPE in this limit.

In the numerical calculations presented in this section and later on,
$n$ was restricted by some finite $N$ ($c_{n} \equiv 0$ for $n>N$).
The cut-off number of atoms $N$ was chosen large enough such that
its influence on the eigenstates is negligible.
For example, for the plots shown in Fig.~\ref{cn}, it was enough to use $N=10$,
but for Figs.~\ref{cn_cs},~\ref{deviation} $N=500$ was used.

\section{\label{E}Excitations}

We consider small perturbation of the ground state
$
c_{{\bf l}n}(t)
=
\left[
    c_{n}^{(0)}
    +
    c_{{\bf l}n}^{(1)}(t)
    +
    \dots
\right]
\exp
\left(
    -i \omega_0 t
\right)
$,
where
\begin{equation}
\label{sol}
c_{{\bf l}n}^{(1)}(t)
=
    u_{{\bf k}n}
    e^{
        i
        \left(
            {\bf k}\cdot{\bf l}
            -
            \omega_{\bf k} t
        \right)
      }
    +
    v_{{\bf k}n}^*
    e^{
        -i
        \left(
            {\bf k}\cdot{\bf l}
            -
            \omega_{\bf k} t
        \right)
      }
\;.
\end{equation}
Substituting this expression into GE and keeping only linear terms
with respect to $u_{{\bf k}n}$ and $v_{{\bf k}n}$, we obtain the system of
linear equations~\cite{TML08}
\begin{equation}
\label{evpexc}
\hbar\omega_{\bf k}
\left(
    \begin{array}{c}
       \vec{u}_{\bf k}\\
       \vec{v}_{\bf k}
    \end{array}
\right)
=
\left(
    \begin{array}{cc}
        A_{\bf k} & B_{\bf k}\\
       -B_{\bf k} & -A_{\bf k}
    \end{array}
\right)
\left(
    \begin{array}{c}
       \vec{u}_{\bf k}\\
       \vec{v}_{\bf k}
    \end{array}
\right)
\;,
\end{equation}
where $\vec{u}_{\bf k}$ and $\vec{v}_{\bf k}$ are infinite-dimensional vectors
with the components $u_{{\bf k}n}$ and $v_{{\bf k}n}$ ($n=0,1,\dots$), respectively.
Matrix elements of $A_{\bf k}$ and $B_{\bf k}$ have the form
\begin{eqnarray}
A_{\bf k}^{nn'}
&=&
-J_{\bf 0}
\psi^{(0)}
\left(
    \sqrt{n'}\,
    \delta_{n',n+1}
    +
    \sqrt{n}\,
    \delta_{n,n'+1}
\right)
\nonumber\\
&+&
\left[
    \frac{U}{2}\,n(n-1)
    -
    \mu n
    -
    \hbar\omega_0
\right]
\delta_{n',n}
\nonumber\\
&-&
J_{\bf k}
\left[
    \sqrt{n+1}\,
    \sqrt{n'+1}\,
    c_{n+1}^{(0)}\,
    c_{n'+1}^{(0)}
\right.
\nonumber\\
    &+&
\left.
    \sqrt{n}\,
    \sqrt{n'}\,
    c_{n-1}^{(0)}\,
    c_{n'-1}^{(0)}
\right]
\;,
\nonumber\\
B_{\bf k}^{nn'}
&=&
-
J_{\bf k}
\left[
    \sqrt{n+1}\,
    \sqrt{n'}\,
    c_{n+1}^{(0)}\,
    c_{n'-1}^{(0)}
\right.
\nonumber\\
    &+&
\left.
    \sqrt{n}\,
    \sqrt{n'+1}\,
    c_{n-1}^{(0)}\,
    c_{n'+1}^{(0)}
\right]
\;,
\nonumber
\end{eqnarray}
where $J_{\bf k}=2dJ-\epsilon_{\bf k}$ with
\begin{equation}
\epsilon_{\bf k}
=
4J
\sum_{\alpha=1}^d
\sin^2
\left(
    \frac{k_\alpha}{2}
\right)
\end{equation}
being the energy of a free particle.
This system is valid for both phases and generalizes the BdG equations
previously derived for coherent states.
The dependence on vector ${\bf k}$ is determined by the variable
\begin{equation}
\label{x}
x
=
\left(
\frac{1}{d}\sum_{\alpha=1}^d \sin^2 \frac{k_\alpha}{2}
\right)^{1/2}
\;,
\end{equation}
which varies from $0$ to $1$.
For small $|{\bf k}|$, $x\approx |{\bf k}|/(2\sqrt{d})$.

The energy increase due to the perturbation is given by~\cite{stringari}
\begin{equation}
\label{dE}
\Delta E
=
\hbar \omega_\vc{k}
\left(
    |\vec{u}_{\vc{k}}|^2-|\vec{v}_{\vc{k}}|^2
\right)
\;.
\end{equation}
Formally, Eqs.~(\ref{evpexc}) have solutions with positive and negative energies
$\pm\hbar\omega_{\bf k}$, which are equivalent because
Eqs.~(\ref{sol}),~(\ref{dE}) are invariant under the transformation
$\omega_{\bf k}\to -\omega_{\bf k}$,
${\bf k}\to -{\bf k}$,
$\vec{u}_{\bf k}\to\vec{v}^*_{\bf k}$,
$\vec{v}^*_{\bf k}\to\vec{u}_{\bf k}$,
so that only solutions with the positive energies will be considered in the following.
The eigenvectors are chosen to follow the orthonormality relations
\begin{equation}
\vec{u}^*_{\vc{k},\lambda'}
\cdot
\vec{u}_{\vc{k},\lambda}
-
\vec{v}^*_{\vc{k},\lambda'}
\cdot
\vec{v}_{\vc{k},\lambda}
=
\delta_{\lambda,\lambda'}
\;.
\nonumber
\end{equation}

Perturbation~(\ref{sol}) creates plane waves of the order parameter
$\psi_{\bf l}(t)=\psi^{(0)}+\psi_{\bf l}^{(1)}(t)$, where
\begin{eqnarray}
\label{psiw}
\psi_{\bf l}^{(1)}(t)
&=&
{\cal U}_{\bf k}
e^{
    i
    \left(
        {\bf k}\cdot{\bf l}
        -
        \omega_{\bf k} t
    \right)
  }
+{\cal V}^*_{\bf k}
e^{
    -i
    \left(
        {\bf k}\cdot{\bf l}
        -
        \omega_{\bf k} t
    \right)
  }
\;,
\\
{\cal U}_{\bf k}
&=&
\sum_n
\left(
    c_{n-1}^{(0)}
    u_{{\bf k}n}
    +
    c_{n}^{(0)}
    v_{{\bf k},n-1}
\right)
\sqrt{n}
\;,
\nonumber\\
{\cal V}_{\bf k}
&=&
\sum_n
\left(
    c_{n}^{(0)}
    u_{{\bf k},n-1}
    +
    c_{n-1}^{(0)}
    v_{{\bf k}n}
\right)
\sqrt{n}
\;.
\nonumber
\end{eqnarray}

The perturbations for the total density and the condensate density are given by
\begin{eqnarray}
\label{dw}
\langle
    \hat n_{\bf l}
\rangle(t)
&=&
\langle
    \hat n
\rangle
+
\left[
{\cal A}_{\bf k}
e^{
    i
    \left(
        {\bf k}\cdot{\bf l}
        -
        \omega_{\bf k} t
    \right)
  }
+
{\rm c.c.}
\right]
\;,
\\
{\cal A}_{\bf k}
&=&
\sum_n
c_{n}^{(0)}
n
\left(
    u_{{\bf k}n}
    +
    v_{{\bf k}n}
\right)
\;,
\nonumber
\end{eqnarray}
and
\begin{eqnarray}
\label{cw}
\left|
    \psi_{\bf l}(t)
\right|^2
&=&
{\psi^{(0)}}^2
+
\left[
{\cal B}_{\bf k}
e^{
    i
    \left(
        {\bf k}\cdot{\bf l}
        -
        \omega_{\bf k} t
    \right)
  }
+
{\rm c.c.}
\right]
\;,
\\
{\cal B}_{\bf k}
&=&
\psi^{(0)}
\sum_n
\left[
    c_{n-1}^{(0)}
    \left(
        u_{{\bf k}n}
        +
        v_{{\bf k}n}
    \right)
\right.
\nonumber\\
    &+&
\left.
    c_{n}^{(0)}
    \left(
        u_{{\bf k},n-1}
        +
        v_{{\bf k},n-1}
    \right)
\right]
\sqrt{n}
\;.
\nonumber
\end{eqnarray}
In what follows we consider the properties of the excitations in the MI and SF phases.
Although the results for the MI are not new, we would like to present those for completeness.

\subsection{Mott insulator}

For the MI phase, the coefficients $c_n^{(0)}$
have a simple analytical form~(\ref{gsmi}).
The eigenvalue problem for the infinite-dimensional matrices~(\ref{evpexc})
reduces to the diagonalization of two $2\times 2$-matrices which couple
$u_{{\bf k},n_0-1}$ to $v_{{\bf k},n_0+1}$ and
$u_{{\bf k},n_0+1}$ to $v_{{\bf k},n_0-1}$, respectively.
The lowest-energy excitation spectrum consists of two branches with the energies
\begin{eqnarray}
\label{om}
\hbar\omega_{{\bf k}\pm}
&=&
\frac{1}{2}
\sqrt{
       U^2
       -
       4J_{\bf k}
       U
       \left(
           n_0+\frac{1}{2}
       \right)
       +
       J_{\bf k}^2
     }
\nonumber\\
&\pm&
\left[U
\left(
    n_0-\frac{1}{2}
\right)
    -\mu
    -\frac{J_{\bf k}}{2}
\right]
\;,
\end{eqnarray}
The same result was obtained using Hubbard-Stratonovich transformation~\cite{Oosten}
and within the Schwinger-boson approach~\cite{Huber}.

\begin{figure}[t]



\hspace{-3cm}
\includegraphics[width=10cm]{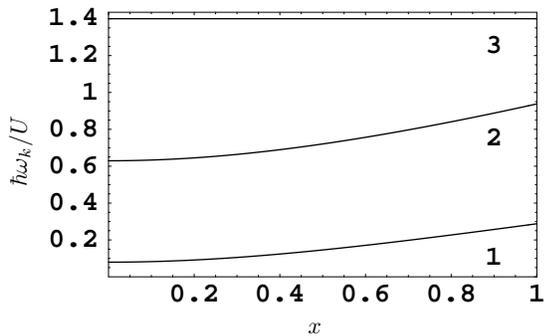}

\caption{
First three branches
$\hbar\omega_{\vc{k}-}$~(1),
$\hbar\omega_{\vc{k}+}$~(2) and
$\hbar\omega_{\lambda=0}$~(3)
of the excitations spectrum of the MI for $\mu/U=1.2$ and $2dJ/U=0.05$,
which corresponds to $n_0=2$.
}
\label{excmi}
\end{figure}

These two branches are shown in Fig.~\ref{excmi} and display a gap.
According to Eqs.~(\ref{cw}),~(\ref{dw}), no density wave is created in the two modes,
although the order parameter does not vanish [see Eq.~(\ref{psiw})].
Eq.~(\ref{om}) can be also rewritten in the form
\begin{eqnarray}
\hbar\omega_{{\bf k}+}
&=&
\epsilon_{{\bf k}p}
- \mu
\;,
\nonumber\\
\hbar\omega_{{\bf k}-}
&=&
-\epsilon_{{\bf k}h}
+
\mu
\nonumber
\;.
\end{eqnarray}
Therefore, the solutions with index '$+$' and '$-$' have the meaning
of the particle and hole excitations, respectively~\cite{EM99}.

Other solutions of Eq.~(\ref{evpexc}) are independent of ${\bf k}$ with the energies
\begin{equation}
\hbar\omega_{\lambda}
=
\frac{U}{2}
\left[
    \lambda(\lambda-1)
    -
    n_0(n_0-1)
\right]
    -
    \mu
    (\lambda-n_0)
\;,
\end{equation}
They are denoted by $\lambda$ which are non-negative integers different from $n_0,n_0\pm 1$.
If $n_0$ is the smallest integer greater than $\mu/U$,
the excitation energies are always positive.
The eigenvectors of these modes have the form
$u_{{\bf k}n\lambda}=\delta_{n,\lambda}$, $v_{{\bf k}n\lambda}=0$,
and the amplitudes of all the waves defined by Eqs.~(\ref{psiw}),~(\ref{dw}),~(\ref{cw})
vanish.

The boundary between the SF and MI phases is determined
from the disappearance of the gap in the excitation spectrum, i.e.,
when $\omega_{\vc{0}-}=0$.
Under this condition, we recover the critical ratio~\cite{Sachdev}:
\begin{equation}
\label{crit}
2d(J/U)_c
=
\frac
{(n_0-\mu/U)(\mu/U-n_0+1)}
{1+\mu/U}
\;.
\end{equation}
It takes its maximal value when
\begin{equation}
\label{Jcmax}
2d(J/U)_c^{max}
=
\left(
    \sqrt{n_0+1}
    -
    \sqrt{n_0}
\right)^2
\end{equation}
for a chemical potential given by
\begin{equation}
(\mu/U)_c=\sqrt{n_0(n_0+1)}-1
\;.
\end{equation}
For $J/U>(J/U)_c$, the lowest frequency $\omega_{\vc{0}-}$ in Eq.~(\ref{om}) 
becomes negative leading to a negative expression for Eq.~(\ref{dE}),
so that the Mott-phase solution (\ref{gsmi}) does not correspond to the ground state anymore.

The excitation spectrum has interesting features on the boundary between the MI and SF.
For $(J/U)_c=(J/U)_c^{max}$, the excitation energies (\ref{om}) can be rewritten as
\begin{equation}
\hbar\omega_{{\bf k}\pm}
=
\left[
    \sqrt{n_0(n_0+1)}
    U
    \epsilon_{\bf k}
    +
    \frac{\epsilon^2_{\bf k}}{4}
\right]^{1/2}
\pm
\frac{\epsilon_{\bf k}}{2}
\;.
\end{equation}
For small $|{\bf k}|$, the two branches are degenerate and have linear dependence
$\omega_{{\bf k}\pm}=c_s^{tip}|{\bf k}|$ with the sound velocity
\begin{equation}
\label{cs0}
c_s^{tip}
=
\frac{U}{\hbar}
\sqrt
{
 (J/U)_c^{max}
}
\left[
    n_0(n_0+1)
\right]^{1/4}
\end{equation}
expressed in the units of number of sites per second.
For other points on the boundary, i.e., $(J/U)_c < (J/U)_c^{max}$,
no degeneracy appears and the sound velocity vanishes leading to
the quadratic dispersion $\omega_{{\bf k}\pm}\sim {\bf k}^2$ for small $|{\bf k}|$.

\subsection{Superfluid}

\begin{figure}[t]



\hspace{-3cm}
\includegraphics[width=10cm]{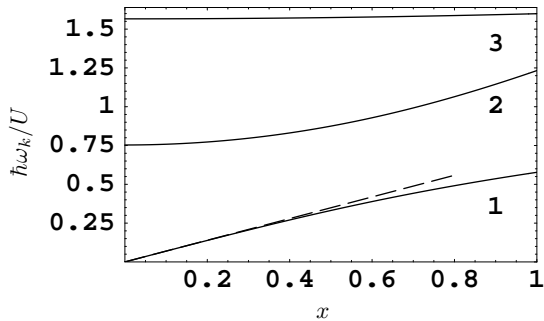}

\caption{
First three branches $\hbar\omega_{\vc{k},\lambda}$ ($\lambda=1,2,3$) of
the excitations spectrum of the SF for $\mu/U=1.2$ and $2dJ/U=0.15$.
The straight dashed line represents the linear approximation
with the sound velocity~(\ref{c_s}).
}
\label{excsf}
\end{figure}

\begin{figure}[t]




\hspace{-3cm}
\includegraphics[width=10cm]{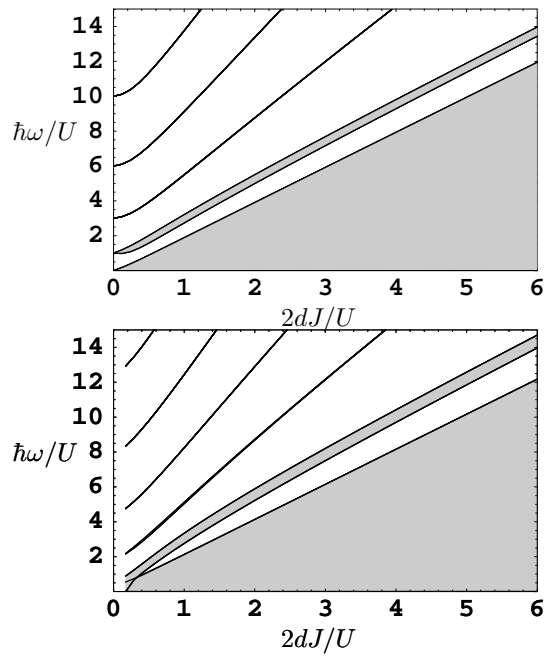}

\caption{
Band structure of the excitation spectrum for
$\langle \hat n\rangle=0.5$~(a), $1$~(b).
Shaded regions which are extremely narrow for higher bands
show allowed excitation energies.
Lower and upper boundaries of the bands correspond to
$x=0$ and $x=1$, respectively.
}
\label{bands_sf}
\end{figure}

\begin{figure}[t]




\hspace{-3cm}
\includegraphics[width=10cm]{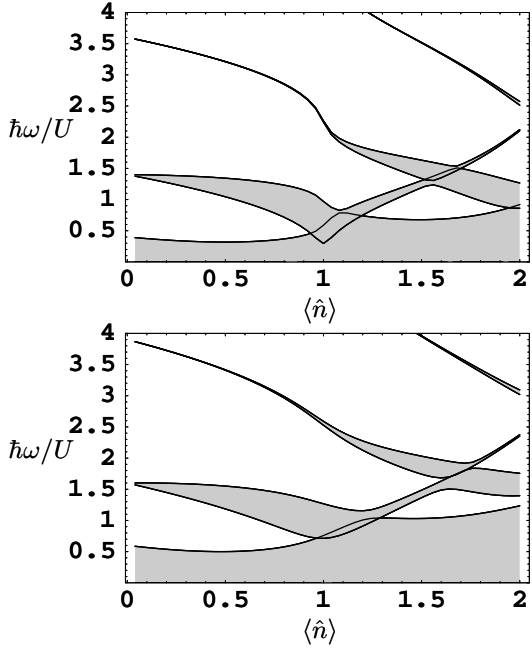}

\caption{
Band structure of the excitation spectrum for
$2dJ/U=0.2$~(a), $0.3$~(b) versus the density.
Shaded regions show allowed excitation energies.
Lower and upper boundaries of the bands correspond to
$x=0$ and $x=1$, respectively.
}
\label{bands_sf1}
\end{figure}

\begin{figure}[t]




\hspace{-3cm}
\includegraphics[width=10cm]{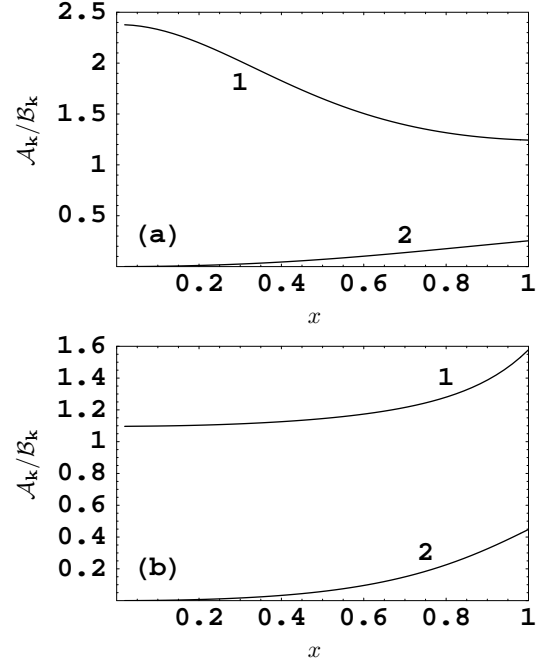}

\caption
{
${\cal A}_{\bf k}/{\cal B}_{\bf k}$ for
$\mu/U=1.2$ and $2dJ/U=0.15$~(a), $1$~(b).
}
\label{AB}
\end{figure}

In the SF phase, the eigenvalue problem~(\ref{evpexc}) is solved using
the numerical values of $c_n^{(0)}$ for each $J/U$ and $\mu/U$.
The energies of the lowest-energy excitations are shown in Fig.~\ref{excsf}.
The excitation spectrum consists of several branches which form a band structure
shown in Figs.~\ref{bands_sf},~\ref{bands_sf1}.
In contrast to the MI, the lowest branch has no gap.
It is a Goldstone mode which appears due to the spontaneous breaking of the phase symmetry
and, therefore, can be called a phase mode~\cite{HTAB}.
As it is shown in Fig.~\ref{AB},
the amplitude of the total-density wave is larger than the amplitude of the condensate-density
wave for this mode. A value greater than unity for the ratio 
${\cal A}_{\bf k}/{\cal B}_{\bf k}$ means that the condensed part and normal part
oscillate in phase.

Higher modes ($\lambda \geq 2$) do not exist in the formalism based on the DGPE.
They have gaps
$\Delta_\lambda=\hbar\omega_{{\bf 0}\lambda}$
which grow with the increase of $J$ (see Fig.~\ref{bands_sf}).
As it is shown in Appendix~\ref{EG}, for large $J$ they have an asymptotic form
\begin{equation}
\label{gap2}
\Delta_\lambda
=
2dJ
\lambda
+
\frac{U}{2}
\lambda
\left(
    \langle \hat n\rangle
    +
    \lambda - 1
\right)
\;.
\end{equation}
We note that only the first two lowest-energy branches have a strong dependence on ${\bf k}$.
For the second mode ($\lambda=2$), the amplitude of the total-density wave is much less than that
of the condensate-density wave (see Fig.~\ref{AB}) which
means that the oscillations of the condensate and normal components are out-of-phase.
However, this does not necessarily mean that there is an
exchange of particles between the condensate and normal component~\cite{Huber,Bissbort}.
Due to the reasons explained in Refs.~\cite{Huber,HTAB,Bissbort}
this type of excitations is called an amplitude mode.
In the rest part of this section, we will study in more details the properties of the Goldstone mode.

As shown in Appendix~\ref{Der}, the lowest-energy branch has a linear form
$\omega_{\vc{k},1}=c_s^0|\vc{k}|$
for small ${\bf k}$ with the sound velocity given by
\begin{equation}
\label{c_s}
c_s^0
=
\sqrt{\frac{2J}{\kappa}}
\left|
    \psi^{(0)}
\right|
/{\hbar}
\;,
\end{equation}
where $\kappa=\frac{\partial\langle\hat n\rangle}{\partial\mu}$ is the compressibility.
This result proves that the Gutzwiller approximation is {\it gapless} and coincides with
Eq.~(\ref{vs}).

\begin{figure}[t]

\hspace{-2.5cm}
\includegraphics[width=10cm]{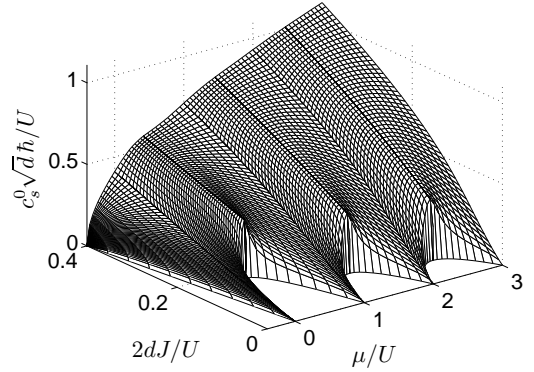}

\caption{
Sound velocity calculated numerically from Eq.~(\ref{c_s}).
Note the discontinuities at the points
$[(J/U)_c^{max},(\mu/U)_c]$
described by Eq.~(\ref{cs0}).
}
\label{sv}
\end{figure}

Fig.~\ref{sv} shows the dependence of the sound velocity on $\mu$ and $J$
calculated numerically using Eq.~(\ref{c_s}). If we approach the boundary of the MI,
the sound velocity goes to zero everywhere except the tips of the lobes,
where it is perfectly described by Eq.~(\ref{cs0}).
This behavior can be understood considering the properties of $\psi^{(0)}$
and $\kappa$. If we approach the SF-MI transition from the SF part of the phase diagram,
the order parameter $\psi^{(0)}$ tends always continuously to zero.
The compressibility $\kappa$ reaches a finite value at every point of the boundary
except the tips of the MI-lobes where it tends continuously to zero such that
the ratio $\psi^{(0)}/\sqrt{\kappa}$ is finite. Therefore, the sound velocity
vanishes at any point of the phase boundary except the tips of the lobes~\cite{Fisher,Menotti08}.

\begin{figure}[t]



\hspace{-3cm}
\includegraphics[width=10cm]{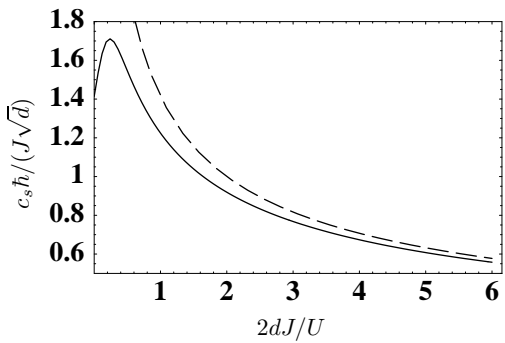}

\caption{Comparison of the sound velocity calculated numerically
from Eq.~(\ref{c_s}) (solid line) with the analytical expression
(\ref{csB}) (dashed line) for
$\langle \hat n\rangle=0.5$.
}
\label{sv2}
\end{figure}

For a weakly interacting gas ($U \ll J$),
$|\psi^{(0)}|^2\approx \langle\hat n\rangle$ and $\kappa\approx 1/U$.
In this limit, we recover the Bogoliubov's dispersion relation (see Appendix~\ref{Bdr})
and the expression for the sound velocity~\cite{Menotti04,Zaremba}
\begin{equation}
\label{csB}
c_s^B
=
\sqrt{2JU\langle\hat n\rangle}/\hbar
\;.
\end{equation}
The comparison of the exact numerical values of the sound velocity calculated
according to Eq.~(\ref{c_s}) with the approximation (\ref{csB}) is shown in Fig.~\ref{sv2}.
As it is expected, the agreement is good at large $J$ but for small tunneling rates
the behavior of $c_s^B$ is completely different.

In the opposite limit ($J \ll U$), the superfluid regions with the atomic densities
$n_0 < \langle\hat n\rangle < n_0+1$ are confined in the regions $\mu_- < \mu < \mu_+$,
where $\mu_-$ is the upper boundary of the MI with $n_0$
and $\mu_+$ is the lower boundary of the MI with $n_0+1$.
For $J \rightarrow 0$, $\mu$ is a linear function of the density, i.e.,
\begin{equation}
\mu
=
\mu_-
+
\left(
    \mu_+ - \mu_-
\right)
\left(
    \langle\hat n\rangle - n_0
\right)
\;.
\end{equation}
Using Eq.~(\ref{crit}) up to the first order in $J$, we obtain
\begin{equation}
\mu_\pm= U n_0 \pm 2dJ(n_0+1)
\end{equation}
from which we deduce
\begin{equation}
\label{kappa}
\kappa=1/[4dJ(n_0+1)]
\;.
\end{equation}
Using $\partial E / \partial \langle\hat n\rangle = \mu$
and the condition that the energy at the phase boundaries
$E_\pm =Un_0(n_0\pm 1)/2$,
the total energy becomes in this limit
\begin{eqnarray}
\label{En}
E
&=&
U
n_0
\left(
    \langle\hat n\rangle
    -
    \frac{n_0+1}{2}
\right)
\\
&-&
2dJ
\left(
    n_0+1
\right)
\left(
    \langle\hat n\rangle-n_0
\right)
\left(
    n_0+1 - \langle\hat n\rangle
\right)
\;.
\nonumber
\end{eqnarray}
In order to deduce the order parameter, we use the relation
$\partial E /\partial J= -2d |\psi^{(0)}|^2$.
From Eq.~(\ref{En}) we obtain
\begin{equation}
\label{psi2}
|\psi^{(0)}|^2
=
(n_0+1)
(\langle \hat n\rangle -n_0)
(n_0+1-\langle n \rangle)
\;.
\end{equation}
Eqs.~(\ref{En}),~(\ref{psi2}) coincide with the ones obtained in Ref.~\cite{Krauth}
but using a perturbation approach.
Substituting (\ref{kappa}) and (\ref{psi2}) into (\ref{c_s}), we obtain finally:
\begin{eqnarray}
\label{csu}
c_s^0
=
2J(n_0+1)
\sqrt{2d(\langle \hat n \rangle-n_0)(n_0+1-\langle \hat n \rangle)}
/\hbar
\;.
\end{eqnarray}
The sound velocity in the limit of small $J$ vanishes at
$\langle\hat n\rangle=n_0,n_0+1$
and takes maximal values at $\langle\hat n\rangle=n_0+1/2$.
This qualitative behavior is the same as
in the case of hard-core bosons in 1D, where the sound velocity is given by
$c_s^{HC}=2J\sin(\pi \langle\hat n\rangle)/\hbar$ for
$0<\langle\hat n\rangle<1$ (see, e.g.,~\cite{Cazalilla}).

Close to the phase boundary, the sound velocity can be calculated analytically
according to Eq.~(\ref{c_s}) within the fourth-order perturbation theory~\cite{Oosten}.
The result for $c_s$ appears to be very long and cannot be displayed here.
However, it describes all the qualitative features discussed above and reproduces
Eq.~(\ref{cs0}).

\section{\label{T}Transition processes}

\subsection{Bragg scattering}

The Bragg scattering is a common experimental method to measure
the excitation spectrum of an ultra-cold gas~\cite{stringari}. This process 
is induced by the following perturbation term in the Hamiltonian:
\begin{eqnarray}
\label{Hint}
\hat H'(t)
&=&
\sum_{\bf l}
V_{\vc{k},\omega}
\cos(\vc{k}\cdot\vc{l}-\omega t)
\hat a^\dagger_{\bf l} \hat a_{\bf l}
\;,
\end{eqnarray}
where $\omega$ and $\vc{k}$ are the frequency and the wavevector of the excitation,
respectively.
This perturbation changes the gas density according to:
\begin{eqnarray}
\delta n_\vc{l}(t)
=
\langle \hat n_\vc{l}(t)\rangle -
\langle \hat n \rangle=\frac{1}{2}
\delta \rho_{\vc{k},\omega}
e^{
    i
    \left(
        {\bf k}\cdot{\bf l}
        -
        \omega_k t
    \right)
  }
+
{\rm c.c.}
\end{eqnarray}
Up to the first order, this change is linear in the Bragg potential, i.e.,
\begin{eqnarray}
\label{drho}
\delta \rho_{\vc{k},\omega}
=
\chi(\vc{k},\omega)V_{\vc{k},\omega}
\;,
\end{eqnarray}
where  $\chi(\vc{k},\omega)$ is the susceptibility.
Within the Gutzwiller approximation, we obtain
\begin{eqnarray}
\label{chi}
&&
\chi(\vc{k},\omega)=
\\
&&
-
\left(
    \begin{array}{c}
       \vec{u}^{(0)}_{\bf 0}\\
       \vec{u}^{(0)}_{\bf 0}
    \end{array}
\right)^T
\left(
    \begin{array}{cc}
        A_{\bf k}-\hbar\tilde\omega & B_{\bf k}\\
        B_{\bf k} & A_{\bf k}+\hbar\tilde\omega
    \end{array}
\right)^{-1}
\left(
    \begin{array}{c}
       \vec{u}^{(0)}_{\bf 0}\\
       \vec{u}^{(0)}_{\bf 0}
    \end{array}
\right)
\;,
\nonumber
\end{eqnarray}
where $\tilde\omega=\omega+i0$
and the vector components are defined as
$u^{(0)}_{{\bf 0}n} = n c^{(0)}_{n}$.
After matrix inversion, the susceptibility can be rewritten in the more simple form
\begin{eqnarray}
\label{chi2}
\chi(\vc{k},\omega)
=
\frac{2}{\hbar}
\sum_\lambda
\frac
{\chi_{\vc{k},\lambda} \omega_{\vc{k},\lambda}}
{
 \left(
     \omega + i0
 \right)^2
 -
 \omega_{\vc{k},\lambda}^2
}
\;,
\end{eqnarray}
where $\lambda$ denotes various excitation branches
associated to the eigenvalues $\omega_{\vc{k},\lambda}$ and
\begin{equation}
\chi_{\vc{k},\lambda}
=
\left|
    \vec{u}^{(0)}_{\bf 0}\cdot(\vec{u}_{\vc{k},\lambda}+\vec{v}_{\vc{k},\lambda})
\right|^2
\end{equation}
is the amplitude for the Bragg scattering of transition frequency
$\omega_{\vc{k},\lambda}$. As we see, $\chi_{\vc{k},\lambda}$ is nothing but
the square of amplitude of the density wave defined by Eq.~(\ref{dw}).

For long wavelength, only the lowest mode is dominant since
$\chi_{\vc{k},\lambda}$ vanishes for the higher modes.
This is a consequence of the orthogonality between the eigenvector components
of the other mode and  $(\vec{u}^{(0)}_{\bf 0}, -\vec{u}^{(0)}_{\bf 0})$
in the long-wavelength limit (see Appendix~\ref{Der}).
Since $\lambda=1$ denotes the sound branch, we obtain the approximate expression
\begin{eqnarray}
\label{chi3}
\chi(\vc{k},\omega)
\stackrel{\vc{k} \rightarrow \vc{0}}{=}
\frac{2}{\hbar}
\frac{\chi_{\vc{k},1} \omega_{\vc{k},1}}{(\omega +i0)^2-\omega^2_{\vc{k},1}}
\;.
\end{eqnarray}
The comparison of (\ref{chi3}) with the identity
\begin{equation}
\kappa
=
-
\chi(\vc{k}=\vc{0},\omega=0)
\;,
\end{equation}
which follows from Eq.~(\ref{drho}),
allows to deduce that for long wavelength
\begin{eqnarray}
\label{chiapp}
\chi_{\vc{k},1}
\stackrel
{
 \vc{k} \rightarrow \vc{0}
}
{=}
\frac{\kappa}{2}
c_s^0
|\vc{k}|
\;.
\end{eqnarray}

\begin{figure}[t]



\hspace{-3cm}
\includegraphics[width=10cm]{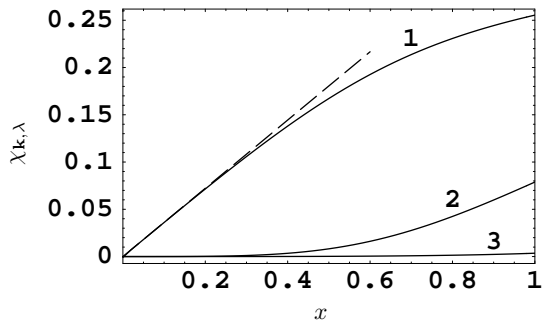}

\caption
{
Transition amplitudes $\chi_{\vc{k},\lambda}$ associated the 
transition frequency $\omega_{\vc{k},\lambda}$ for the lowest excitation
branches ($\lambda=1,2,3$) and
for $\mu/U=1.2$ and $2dJ/U=0.15$.
The dashed line corresponds to the approximation~(\ref{chiapp}).
}

\label{chisf}
\end{figure}

The dependences of $\chi_{{\bf k},\lambda}$ on the variable $x$ defined by Eq.~(\ref{x})
for the excitation branches
with $\lambda=1,2,3$ are shown in Fig.~\ref{chisf}. For the chosen values of parameters,
only two lowest branches display noticeable amplitudes.
Similar results have been also obtained in Ref.~\cite{Huber}.
However, the calculations in Ref.~\cite{Huber} are valid only close to the boundaries MI-SF
because the occupation numbers $n$ in Eq.~(\ref{state}) were restricted to $n=n_0,n_0\pm 1$.

\begin{figure}[t]




\hspace{-3cm}
\includegraphics[width=10cm]{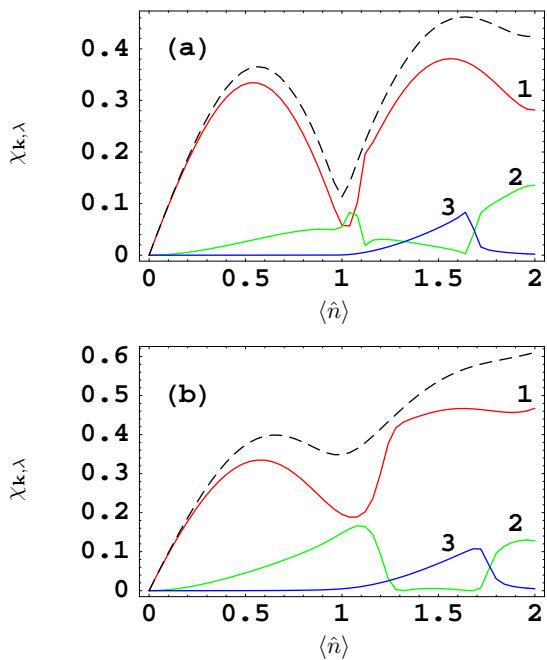}

\caption{
(color online)
Transition amplitudes $\chi_{\vc{k},\lambda}$ versus the density for the first excitation
branches ($\lambda=1,2,3$) and
for $x=1$ and $2dJ/U=0.2$~(a), $0.3$~(b).
Dashed lines show the static structure factor $S({\bf k})$.
}
\label{chisf1}
\end{figure}

In Fig.~\ref{chisf1} instead, we see that the amplitude for the third excitation branch
as well as for the second one can become significant at certain densities.
We would like to note that the f-sum rule is automatically fulfilled in our approach
(see Sect.~\ref{Sr}) in contrast to Refs.~\cite{Oosten2,Huber}.

In the MI phase, the Gutzwiller approximation does not allow to observe any
branches since $\chi_{\vc{k},\lambda}\equiv 0$. No Bragg response is possible,
although the excitations exist in the mean-field approach.
In order to allow non-vanishing response, correlations between
different sites should be included which goes beyond the Gutzwiller approximation~\cite{Navez,Huber}.
In such a description, excitations in the Bragg process are created
as particle-hole pairs~\cite{Huber,Oosten2,Navez}.
However as pointed out in \cite{Navez},
this last process appears to be of the second order in
the inverse of coordination number $z=2d$ and, therefore, is not taken into account
by the standard Gutzwiller ansatz.



\subsection{One-particle Green's function}

The one-particle Green's function can be also determined in the context
of the Gutzwiller approximation through the interaction term
\begin{eqnarray}
\label{Hint2}
\hat H'(t)
&=&
\sum_{\bf l} \eta_{\vc{k},\omega}
e^{i(\vc{k}\cdot\vc{l}-\omega t)}
\hat a^\dagger_{\bf l}
+
{\rm h.c.}
\;,
\end{eqnarray}
which explicitly breaks the $U(1)$ symmetry.
This interaction term induces a deviation in the order parameter
\begin{eqnarray}
\left( \begin{array}{c}
\delta \psi_{\bf l}
\\
\delta \psi^*_{\bf l}
\end{array}\right)
=
\left( \begin{array}{c}
\psi_{\bf l}- \psi^{(0)}
\\
\psi^*_{\bf l}- (\psi^{(0)})^*
\end{array}\right)
\nonumber \\
=
\underline{\underline {G}}(\vc{k},\omega).
\left( \begin{array}{c}
\eta_{\vc{k},\omega}
e^{i(\vc{k}\cdot\vc{l}-\omega t)}
\\
\eta^*_{\vc{k},\omega}
e^{-i(\vc{k}\cdot\vc{l}-\omega t)}
\end{array}\right)
\end{eqnarray}
The proportionality term is the one-particle $2\times2$ 
matrix Green's function with the general expression
\begin{eqnarray}
\label{G}
\underline{\underline{G}}(\vc{k},\omega)
=
\sum_\lambda
\frac{\underline{\underline{g}}_{\vc{k},\lambda}}
{\omega +i0 -  \omega_{\vc{k},\lambda}}
\;,
\end{eqnarray}
where we define the matrix transition amplitude as:
\begin{equation}
\underline{\underline{g}}_{\vc{k},\lambda}
= \underline{b}_{\vc{k},\lambda} .\underline{b}^\dagger_{\vc{k},\lambda}
\end{equation}
and
\begin{equation}
\underline{b}_{\vc{k},\lambda}=\left(
\begin{array}{c}
\sum_{n=1}^\infty
    \left(
       \sqrt{n+1} u_{{\vc{k}},n+1,\lambda}+ \sqrt{n}v_{{\vc{k}},n-1, \lambda}
    \right)
    c^{(0)}_{n}
\\
\sum_{n=1}^\infty
\left(
\sqrt{n+1} v_{{\vc{k}},n+1,\lambda}+ \sqrt{n}u_{{\vc{k}},n-1, \lambda}
\right) c^{(0)}_{n}
\end{array}
\right)
\end{equation}
Here, $\lambda$ denotes branches with both positive and negative energies.
In the SF phase,
the existence of the order parameter {\it hybridizes}
the one-  and the two-particle
Green's functions so that their poles are identical.
On the other hand, in the MI phase, we note that the transitions forbidden
in the Bragg scattering become allowed in the interaction term
Eq.~(\ref{Hint2}).
Indeed, the time-dependent Gutzwiller approach allows to recover the results previously
established in the context of quantum field theory~\cite{Oosten2}:
\begin{eqnarray}
\label{G2}
\underline{\underline{G}}(\vc{k},\omega)
=\underline{\underline{1}}
\sum_\pm
\frac{g_{\vc{k},\pm}}{\omega +i0 \mp \omega_{\vc{k},\pm}}
\;,
\end{eqnarray}
where
\begin{eqnarray}
g_{\vc{k},\pm}
=
\frac{1}{2}
\pm
\frac{(2n_0+1)U- J_\vc{k}}{\hbar (\omega_{\vc{k},+} +\omega_{\vc{k},-})}
\end{eqnarray}
is the probability to create a particle (hole) excitation. 
Although the one-particle Green's function is a concept
of importance in the context of quantum field theory, its use in the concrete
experiments is limited by the impossibility to create the $U(1)$ 
symmetry breaking interaction
(\ref{Hint2}). However, this function  helps to interpret 
the nature of the excitation  which is particle-like when one atom
is added to the gas or hole-like when one atom is removed.

\subsection{\label{Sr}Sum rules}

Let us examine some sum rules satisfied by the susceptibility function.
The compressibility sum rule is deduced from the Kramers-Kronig relation
\begin{eqnarray}
\label{comp}
\int_{-\infty}^\infty \frac{d\omega}{\pi\omega}
\chi''(\vc{k},\omega) = \chi(\vc{k},0)
\;.
\end{eqnarray}
The $f$-sum rule generalizes
the one obtained for a Bose gas in continuum~\cite{Huber}
\begin{eqnarray}
\hbar^2
\int_{-\infty}^\infty
\frac{d\omega}{2\pi}
\omega \chi''(\vc{k},\omega)
=
\nonumber\\
\label{fsum}
-
\sum_{\alpha=1}^d
\int
d^3\vc{k}'
\cos(k_\alpha')
\langle
    \hat a^\dagger_{\vc{k}'}
    \hat a_{\vc{k}'}
\rangle
4J\sin^2(k_\alpha/2)
\;.
\end{eqnarray}
In order to recover the continuum limit, we have to introduce the lattice
constant $a$ by means of the replacement ${\bf k}\to a{\bf k}$.
In the limit of small $a$, we get~\cite{stringari}
\begin{eqnarray}
\int_{-\infty}^\infty
\frac{d\omega}{2\pi}
\omega
\chi''(\vc{k},\omega)
=
-\langle \hat n \rangle \frac{\vc{k}^2}{2m}
\;,
\end{eqnarray}
where $m=\hbar^2/(2Ja^2)$ corresponds to the effective mass.
The application of Eq.~(\ref{comp}) and Eq.~(\ref{fsum}) in the Gutzwiller approximation leads to
\begin{eqnarray}
\int_{-\infty}^\infty \frac{d\omega}{\pi\omega}
\chi''(\vc{k},\omega) &\stackrel{\vc{k}\rightarrow 0}{=}& -\kappa
\;,
\\
\int_{-\infty}^\infty 
\frac{d\omega}{2\pi} \omega \chi''(\vc{k},\omega) &=& 
-
\left|
    \psi^{(0)}
\right|^2
\epsilon_\vc{k}
\;.
\end{eqnarray}
Using the result~(\ref{chi2}), we obtain
\begin{eqnarray}
\sum_\lambda
\frac{\chi_{\vc{k},\lambda}}{\omega_{\vc{k},\lambda}}
&\stackrel{\vc{k}\rightarrow 0}{=}&
\frac{\kappa}{2}
\;,
\\
\hbar^2
\sum_\lambda
\chi_{\vc{k},\lambda}
\omega_{\vc{k},\lambda}
&=&
|\psi^{(0)}|^2
\epsilon_\vc{k}
\;.
\end{eqnarray}
A third sum rule concerns the static structure factor $S(\vc{k})$.
Using the fluctuations dissipation theorem, the dynamic structure factor is expressed
as $S(\vc{k},\omega)=\chi''(\vc{k},\omega)/\pi$ at zero temperature so that 
\cite{stringari}:
\begin{eqnarray}
\label{sfactor}
\int_{0}^\infty \frac{d\omega}{\pi}
\chi''(\vc{k},\omega) = S(\vc{k})
=
\langle
    \delta \hat \rho_\vc{k}
    \delta \hat \rho_\vc{-k}
\rangle
\;.
\end{eqnarray}
where $\delta \hat \rho_\vc{k}=\sum_\vc{l} 
\delta \hat n_\vc{l} e^{-i\vc{k}.\vc{l}}/L^{d/2}$.
This third sum rule is not fulfilled in the Gutzwiller approximation because
the correlation function is equal to the particle number fluctuations
$
\langle
    \delta \hat \rho_\vc{k}
    \delta \hat \rho_\vc{-k}
\rangle
=
\langle
\delta^2 \hat n
\rangle
$
and thus has no $\vc{k}$-dependence.
However, as in the case of a Bose gas in continuum,
this sum rule allows to deduce the static structure factor.
We find indeed
\begin{eqnarray}
S(\vc{k})
=
\sum_\lambda \chi_{\vc{k},\lambda}
\stackrel{\vc{k} \rightarrow \vc{0}}
{=}
\frac{\kappa}{2} c_s |\vc{k}|
\;.
\end{eqnarray}
This last result shows an interesting feature of the sum rule approach.
Starting from the lowest-order Gutzwiller approach that does not contain any
correlation, the two-point correlation function is determined as a next order
contribution. Similarly, starting from the time-dependent DGPE,
a similar procedure has been successfully used to recover the static structure
predicted from the Bogoliubov approach~\cite{stringari,Nozieres}.

\subsection{Spectra measurement}

The observation of the excitation branches in the SF phase can be made through
the measurement of the total momentum ${\bf P}$.
After an adequate time of flight $t$, the momentum is given by~\cite{Dalibard}
\begin{eqnarray}
\vc{P}
=
\int_V d^3 \vc{x}
\frac{M\vc{x}}{t}
n(\vc{x})
=
\hbar
\int d^3 \vc{k} \vc{k} |w(\vc{k})|^2
G(\vc{k})
\;,
\end{eqnarray}
where $w(\vc{p})$ is the Fourier transform of the Wannier function and
\begin{eqnarray}
G(\vc{k})
=
\sum_{\vc{l},\vc{l'}}
e^{i\vc{k}\cdot(\vc{l}-\vc{l'})}
\langle
    \hat a^\dagger_{\bf l}
    \hat a_{\bf l'}
\rangle
\;.
\end{eqnarray}
For small momentum, we can assume $w(\vc{k})\simeq  w(\vc{0})$.
Calculations up to the second order in the potential allow to deduce
\begin{eqnarray}
\frac{d \vc{P}}{dt}
&=&
-2\vc{k}|w(\vc{0})|^2
|\frac{V_{\vc{k},\omega}}{2}|^2
{\rm Im} \chi(\vc{k},\omega)
\\
&=&
2\pi
\vc{k}
|w(\vc{0})\frac{V_{\vc{k},\omega}}{2}|^2
\sum_{\pm,\lambda}
\pm
\chi_{\vc{k},\lambda}
\delta(\omega \mp  \omega_{\vc{k},\lambda})
\;.
\nonumber
\end{eqnarray}

\section{\label{SW}Creation of sound waves}

Experimentally sound waves in a trapped Bose-Einstein condensate were created
turning on and off a perturbation potential in the center of
the atomic cloud~\cite{Ketterle97,MKS09}.
The same can be done in optical lattices and numerical simulations
of this kind of experiment were performed in Ref.~\cite{Menotti04}
deep in the SF phase making use of the DGPE and in Ref.~\cite{KSDZ05}
for 1D-systems using DMRG method.
In this section, we do the same simulations but using the dynamical Gutzwiller ansatz.
Our aim is to compare the results with that obtained by the other methods and to
extract from the simulations the values of the sound velocity compatible
with that calculated in Sec.~\ref{E}.

We are interested in the solutions for $d$-dimensional lattices
which have a position dependence only in one chosen spatial dimension.
Then in Eq.~(\ref{GEd})
$\psi_{{\bf l}\pm{\bf e}_\alpha} \equiv \psi_{l\pm 1}$, if $\alpha$ is the chosen dimension,
otherwise $\psi_{{\bf l}\pm{\bf e}_\alpha} \equiv \psi_{l}$.
Here $l$ is the site index along the chosen dimension.
Initially the atoms are prepared in the ground state of the external potential
\begin{equation}
\label{potential}
\varepsilon_l
=
\varepsilon_0
\exp
\left[
    -(l-l_0)^2/w^2
\right]
\;,\quad
l_0=(L+1)/2
\;,
\end{equation}
where $\varepsilon_0$ and $w$ are the strength and the spatial width of the potential, respectively.
The external potential (\ref{potential}) creates a density perturbation
and we calculate the corresponding ground state numerically propagating Eq.~(\ref{GEd})
in imaginary time~\cite{Dalfovo} with the initial conditions
$c_{ln}(0)=c_{ln}^{(0)}$, where $c_{ln}^{(0)}$ are the coefficients for the ground state
of the homogeneous lattice with the local chemical potential $\mu_l=\mu-\varepsilon_l$.
At $t=0$, the external potential (\ref{potential}) is switched off ($\varepsilon_l\equiv 0$)
and the ground state starts to evolve.
We calculate numerically the evolution of the ground state in real time.
Numerical calculations presented in this section are performed for $d=3$
and we used $L=200$, $N=10$,
which is enough to avoid parasitic effects due to the reflection from the boundaries
and due to the cut-off in the occupation numbers.

\subsection{Deep in the SF phase}

\begin{figure}[t]




\hspace{-3cm}
\includegraphics[width=10cm]{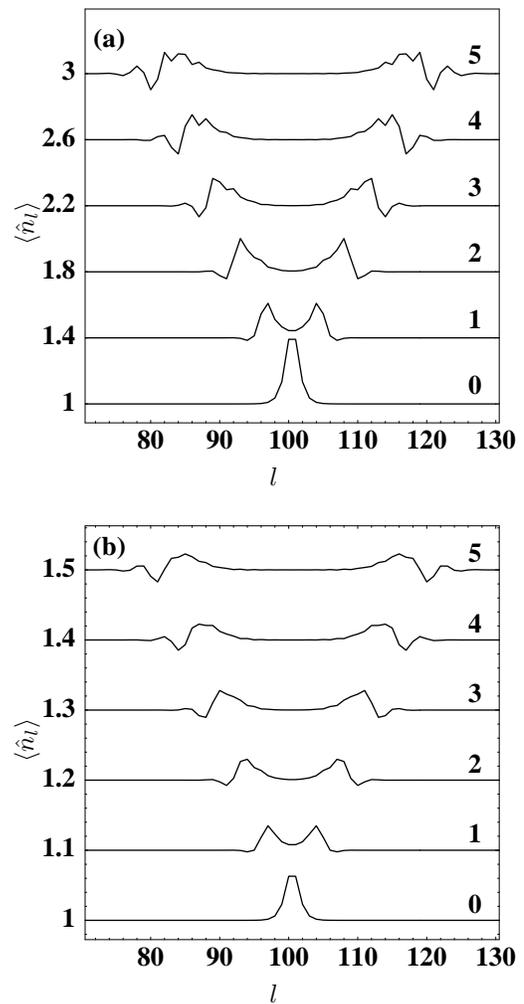}

\caption{
Time evolution of the mean occupation numbers $\langle\hat{n}_l\rangle$
deep in the SF phase after switching off the potential with $w=1$,
$\varepsilon_0/U=-0.6$~(a), $-0.1$~(b).
The parameters are $\langle \hat n \rangle=1$, $2dJ/U=6$.
The curves show the spatial dependences at the dimensionless time $\tau=tU/\hbar=$
$0$~(0), $2$~(1), $4$~(2), $6$~(3), $8$~(4), $10$~(5).
They are shifted by $0.4$~(a) and $0.1$~(b) in the vertical direction
with respect to the previous one.
}
\label{nav-J_6-n_1-w_1}
\end{figure}

\begin{figure}[t]



\hspace{-3cm}
\includegraphics[width=10cm]{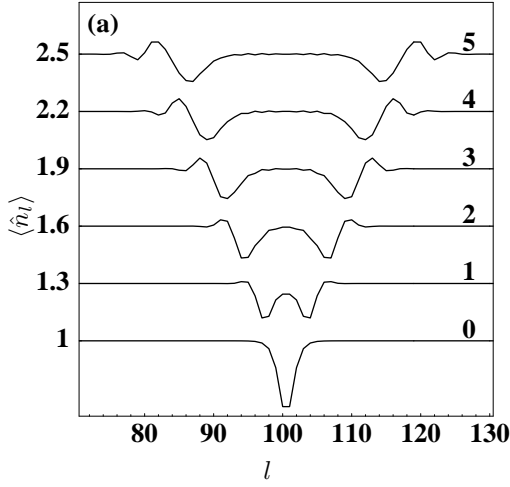}

\caption{
Time evolution of the mean occupation numbers $\langle\hat{n}_l\rangle$
deep in the SF phase after switching off the potential with $w=1$,
$\varepsilon_0/U=0.6$.
The parameters are $\langle \hat n \rangle=1$, $2dJ/U=6$.
The curves show the spatial dependences at the dimensionless time $\tau=tU/\hbar=$
$0$~(0), $2$~(1), $4$~(2), $6$~(3), $8$~(4), $10$~(5).
They are shifted by $0.3$ in the vertical direction
with respect to the previous one.
}
\label{nav-J_6-n_1-a_0.6-w_1}
\end{figure}

First we do simulations deep in the SF phase.
In this case there is no difference between
$\langle\hat n_l\rangle$ and $\left|\psi_l\right|^2$.
Time evolution of $\langle\hat n_l\rangle$ for two negative values of
$\varepsilon_0$ is shown in Fig.~\ref{nav-J_6-n_1-w_1}.
Initially the distribution of the atoms
have a maximum at the position of the potential (bright perturbation).
After switching off the potential the density perturbation splits up in
two wave packets propagating symmetrically outward.
At longer times, the form of the propagating maxima becomes irregular which signals
the creation of shock waves~\cite{Menotti04,KSDZ05}.
The irregularities become more pronounced for larger values of $\left|\varepsilon_0\right|$.
An additional dip arises at the fronts of the wave packets which might stem
from the discreteness of the lattice~\cite{KSDZ05}.
For positive values of $\varepsilon_0$ (gray perturbation),
the dynamics is pretty much the same except that
the maxima are replaced by minima and vice versa and the distribution of the atoms
is more regular (Fig.~\ref{nav-J_6-n_1-a_0.6-w_1}).

\begin{figure}[t]



\hspace{-3cm}
\includegraphics[width=10cm]{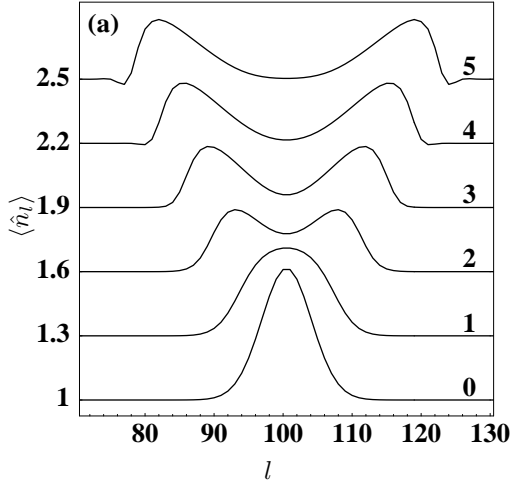}

\caption{
Time evolution of the mean occupation numbers $\langle\hat{n}_l\rangle$
deep in the SF phase after switching off the potential with $w=5$,
$\varepsilon_0/U=-0.6$.
The parameters are $\langle \hat n \rangle=1$, $2dJ/U=6$.
The curves show the spatial dependences at the dimensionless time $\tau=tU/\hbar=$
$0$~(0), $2$~(1), $4$~(2), $6$~(3), $8$~(4), $10$~(5).
They are shifted by $0.3$ in the vertical direction with respect to the previous one.
}
\label{nav-J_6-n_1-a_-0.6-w_5}
\end{figure}

If we increase the width of the external potential $w$, the atomic distribution becomes
more regular (compare Figs.~\ref{nav-J_6-n_1-w_1}(a) and \ref{nav-J_6-n_1-a_-0.6-w_5}).
In Fig.~\ref{nav-J_6-n_1-a_-0.6-w_5} one can clearly see that the form of
the wave packets become very asymmetric during their propagation.

\begin{figure}[t]



\hspace{-3cm}
\includegraphics[width=10cm]{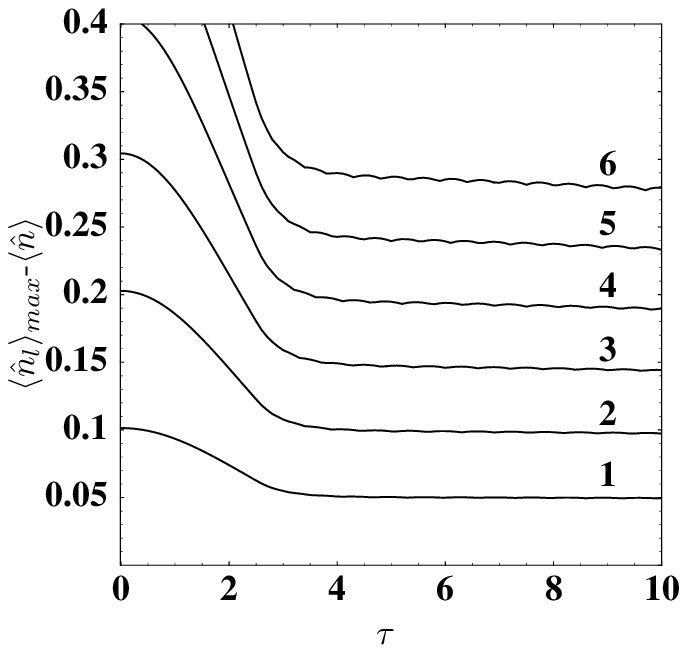}

\caption{
Time evolution of the largest mean occupation number $\langle\hat{n}_l\rangle_{max}$
after switching off the potential with $w=5$,
$\varepsilon_0/U=-0.1$~(1), $-0.2$~(2), $-0.3$~(3), $-0.4$~(4), $-0.5$~(5), $-0.6$~(6).
The parameters are $\langle \hat n \rangle=1$, $2dJ/U=6$.
}
\label{nmax-J_6-n_1-w_5}
\end{figure}

\begin{figure}[t]



\hspace{-3cm}
\includegraphics[width=10cm]{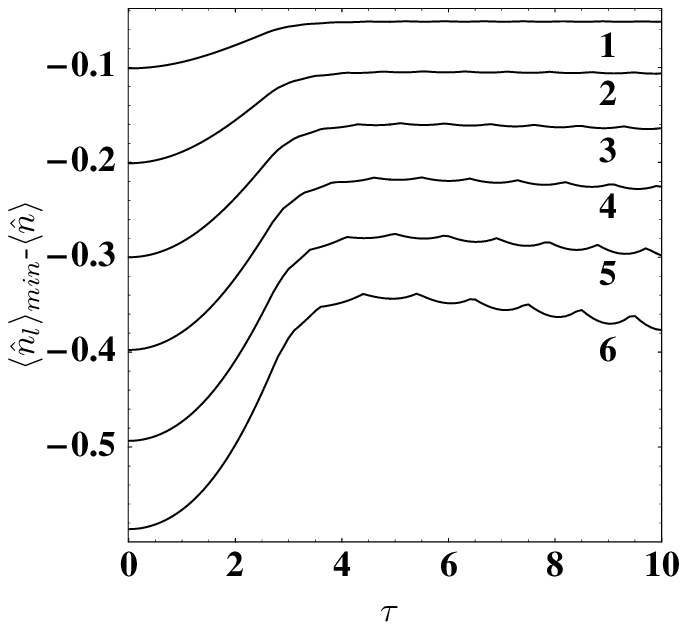}

\caption{
Time evolution of the smallest mean occupation number $\langle\hat{n}_l\rangle_{min}$
after switching off the potential with $w=5$,
$\varepsilon_0/U=0.1$~(1), $0.2$~(2), $0.3$~(3), $0.4$~(4), $0.5$~(5), $0.6$~(6).
The parameters are $\langle \hat n \rangle=1$, $2dJ/U=6$.
}
\label{nmin-J_6-n_1-w_5}
\end{figure}

\begin{figure}[t]



\hspace{-3cm}
\includegraphics[width=10cm]{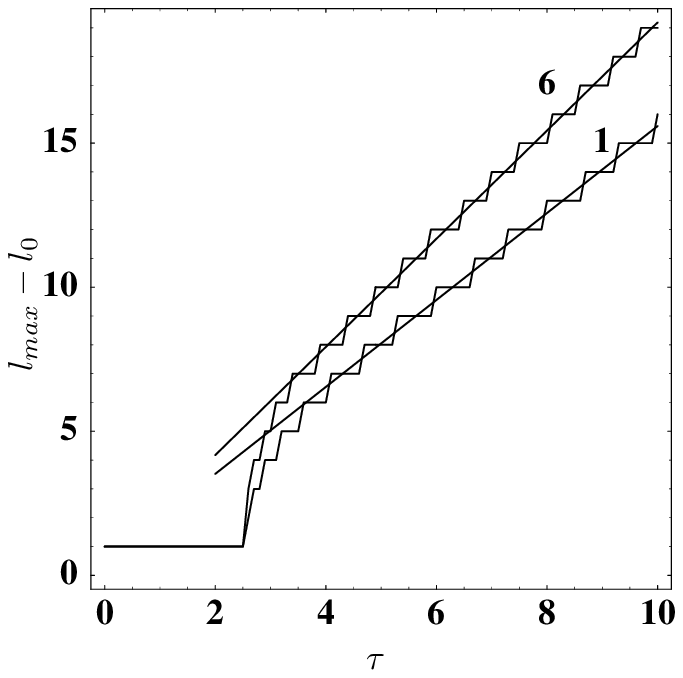}

\caption{
Location of the largest mean occupation number $\langle\hat{n}_l\rangle_{max}$
after switching off the potential with $w=5$,
$\varepsilon_0/U=-0.1$~(1), $-0.6$~(6).
The parameters are $\langle \hat n \rangle=1$, $2dJ/U=6$.
Here we use the same labels for the curves as in Fig.~\ref{nmax-J_6-n_1-w_5}.
Straight lines are linear fits to the numerical data.
}
\label{imax-J_6-n_1-w_5}
\end{figure}

Figs.~\ref{nmax-J_6-n_1-w_5} and \ref{nmin-J_6-n_1-w_5} show the time dependence of the global
maximum and minimum of the atomic distribution $\langle\hat n_l\rangle$ for negative and positive
values of $\varepsilon_0$.
When the external potential is switched off, the amplitude of the density perturbation
goes down and after some finite time seen as the first plateau in Fig.~\ref{imax-J_6-n_1-w_5}
two separate wave packets are formed which propagate in opposite directions.
Their amplitude decreases monotonically in time for negative $\varepsilon_0$
(Fig.~\ref{nmax-J_6-n_1-w_5}) and for small enough positive $\varepsilon_0$
(Fig.~\ref{nmin-J_6-n_1-w_5}). For larger positive $\varepsilon_0$, the amplitude
of the minima increases before it starts to decrease.
Due to the discreteness of the lattice the position of the propagating extremuma
is a step-function of time (Fig.~\ref{imax-J_6-n_1-w_5}) and the amplitude of
the density perturbation shows up oscillations (Figs.~\ref{nmax-J_6-n_1-w_5},~\ref{nmin-J_6-n_1-w_5})
which are stronger for larger values of $\left|\varepsilon_0\right|$.

\begin{figure}[t]



\hspace{-3cm}
\includegraphics[width=10cm]{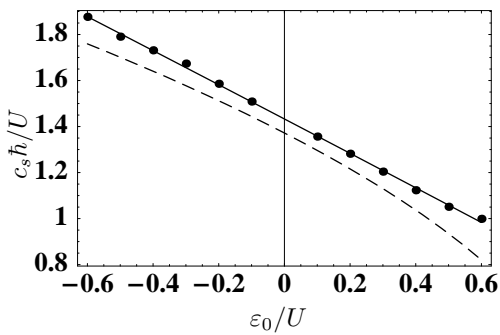}

\caption{
Dependence of the propagation velocity on the strength of the external potential.
The parameters are $\langle \hat n \rangle=1$, $2dJ/U=6$, $w=5$.
The dots are the results of numerical calculations and the solid line is a fit
by quadratic polynomial.
The dashed line shows $c_s^0(\mu'=\mu-\varepsilon_0)$ as a function of $\varepsilon_0$
with $\mu$ fixed by the values of
$\langle\hat n\rangle$ and $2dJ/U$.
}
\label{cs-J_6-n_1-w_5}
\end{figure}

Before the system enters the shock wave regime, there are always pronounced
global maxima (minima) in the case of negative (positive) $\varepsilon_0$
and the propagation velocity of the sound wave packets can be identified
with the velocity of the global extremum. Numerical values of
the propagation velocity $c_s$ in our simulations are determined with the aid of
a linear fit for the location of the global extremum as a function of time
(straight lines in Fig.~\ref{imax-J_6-n_1-w_5}). Its dependence on the amplitude
of the external potential is shown in Fig.~\ref{cs-J_6-n_1-w_5}.
Propagation velocity decreases monotonically with $\varepsilon_0$.
This can be understood looking at the behavior of the function $c_s^0(\mu'=\mu-\varepsilon_0)$
under variation of $\varepsilon_0$ at fixed $2dJ/U$.
For large values of $2dJ/U$, it is also a decreasing function of $\varepsilon_0$
(see Fig.~\ref{sv} and the dashed line in Fig.~\ref{cs-J_6-n_1-w_5}),
which is quite close to the data of our numerical simulations.

\begin{figure}[t]



\hspace{-3cm}
\includegraphics[width=10cm]{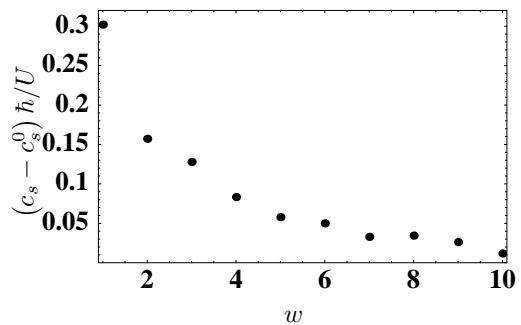}

\caption{
Dependence of the propagation velocity on the width of the external potential.
The parameters are $\langle \hat n \rangle=1$, $2dJ/U=6$.
}
\label{cs-J_6-n_1}
\end{figure}

In order to extract the value of the sound velocity from the numerical data,
we have to extrapolate to $\varepsilon_0=0$. This is done making a quadratic
fit to the data points which is justified by the fact that, near
$\varepsilon_0\approx 0$, $c_s^0(\mu'=\mu-\varepsilon_0)$ can be decomposed into a series
in powers of $\varepsilon_0$.
In the example shown in Fig.~\ref{cs-J_6-n_1-w_5}, the extrapolated value of
the propagation velocity is $1.433$ which is a bit higher than
$c_s^0=1.372$ predicted by the linear response theory.
We have done the same calculations for different values
of $w$ and found that the propagation velocity becomes more close to $c_s^0$
for larger values of $w$ (see Fig.~\ref{cs-J_6-n_1}). Therefore, the deviation
from Eq.~(\ref{c_s}) is due to the contribution of excitations with finite wavelengths.

\subsection{Near the boundary of the SF-MI transition}

\begin{figure}[t]





\hspace{-3cm}
\includegraphics[width=10cm]{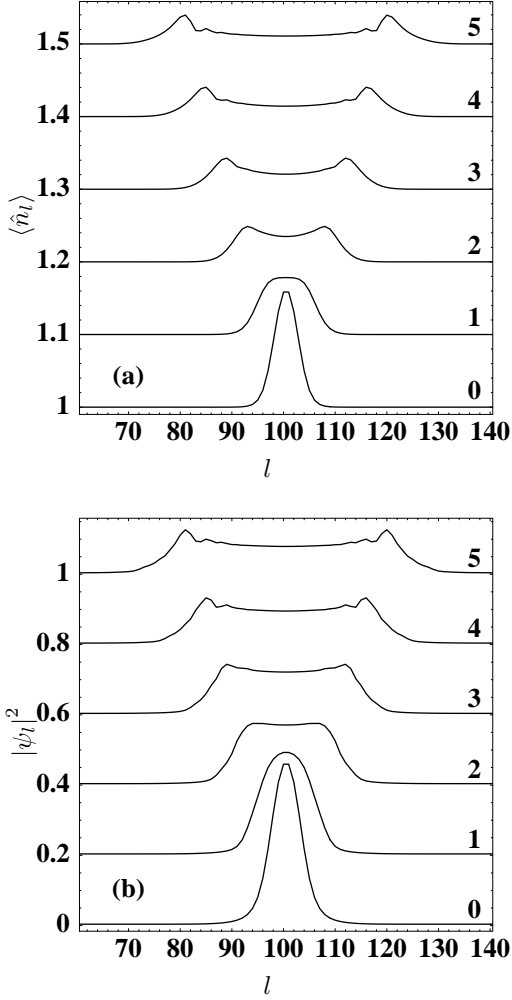}

\caption{
Time evolution of the mean occupation numbers $\langle\hat{n}_l\rangle$~(a)
and the mean number of condensed atoms $|\psi_l|^2$~(b) near the tip of the MI-lobe
after switching off the potential with $\varepsilon_0/U=-0.3$, $w=5$.
The parameters are $\langle \hat n \rangle=1$, $2dJ/U=0.172$.
The curves show the spatial dependences at the dimensionless time $\tau=tU/\hbar=$
$0$~(0), $24$~(1), $48$~(2), $72$~(3), $96$~(4), $120$~(5).
They are shifted by $0.1$~(a) and $0.2$~(b)
in the vertical direction with respect to the previous one.
}
\label{J_0.172-n_1-a_-0.3-w_5}
\end{figure}

Still in the SF phase but near the boundary of the SF-MI transition,
$\left|\psi_l\right|^2 < \langle\hat n_l\rangle$ (Fig~\ref{J_0.172-n_1-a_-0.3-w_5}).
Numerical simulations in this regime show that,
in order to excite only the lowest mode, much less values of $\varepsilon_0$
are required. This is consistent with the fact that the gap between the first
and second excitation modes is very small near the MI lobe.

\begin{figure}[t]



\hspace{-3cm}
\includegraphics[width=10cm]{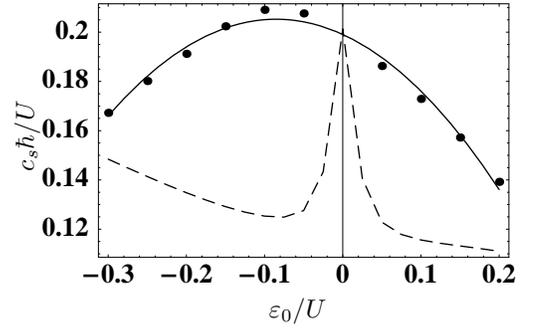}

\caption{
Dependence of the propagation velocity on the strength of the external potential.
The parameters are $\langle \hat n \rangle=1$, $2dJ/U=0.172$, which is
close to the tip of the MI lobe [$2d(J/U)_c^{max}=0.17157$], $w=5$.
The dots are the results of numerical calculations and the solid line is a fit
by quadratic polynomial.
The dashed line shows $c_s^0(\mu'=\mu-\varepsilon_0)$ as a function of $\varepsilon_0$
with $\mu$ fixed by the values of
$\langle\hat n\rangle$ and $2dJ/U$.
}
\label{cs-J_0.172-n_1-w_5}
\end{figure}

Fig.~\ref{cs-J_0.172-n_1-w_5} shows the dependence of the propagation velocity
on $\varepsilon_0$ near the tip of the MI lobe with $n_0=1$.
It is quite different compared to the behavior deep in the SF regime 
where the propagation velocity monotonically decreases (Fig.~\ref{cs-J_6-n_1-w_5}).
Near the tip of the MI lobe, the propagation velocity has a maximum around
$\varepsilon_0\approx 0$ which is qualitatively similar to the behavior of
$c_s^0(\mu'=\mu-\varepsilon_0)$. However, the discrepancy between the propagation velocity
and $c_s^0(\mu'=\mu-\varepsilon_0)$ is large even for small $|\varepsilon_0|$
due to the significant contribution of nonlinear effects.

We have determined the propagation velocity in the limit $\varepsilon_0\to 0$
making again a quadratic fit to the numerical data presented
in Fig.~\ref{cs-J_0.172-n_1-w_5}.
This procedure gives us the value $0.199$, while from Eq.~(\ref{c_s}) we get $c_s^0=0.201$.

\subsection{MI phase}

\begin{figure}[t]





\hspace{-3cm}
\includegraphics[width=10cm]{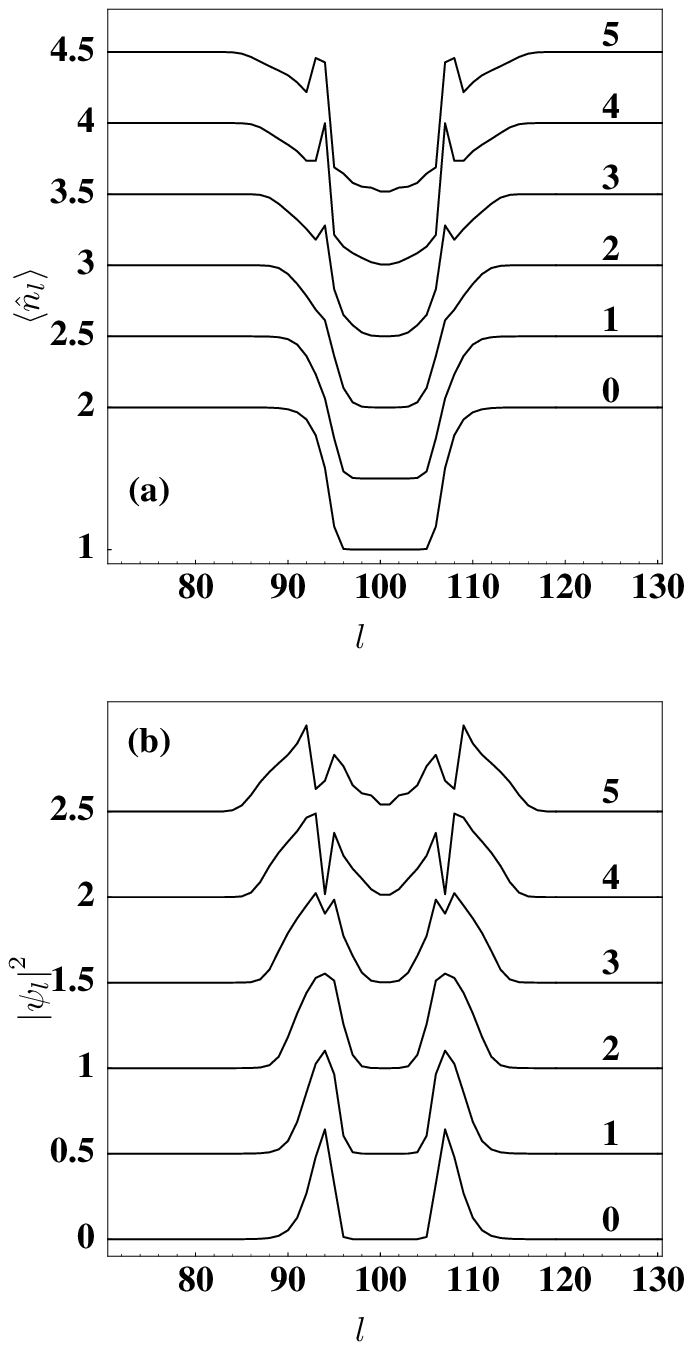}

\caption{
Time evolution of the mean occupation numbers $\langle\hat{n}_l\rangle$~(a)
and the mean number of condensed atoms $|\psi_l|^2$~(b) in the MI phase
after switching off the potential with $\varepsilon_0/U=1$, $w=5$.
The parameters are $\mu/U=1.2$ and $2dJ/U=0.07$, which is a bit smaller than
the critical value $2d(J/U)_c=0.073$.
The curves show the spatial dependences at the dimensionless time $\tau=tU/\hbar=$
$0$~(0), $20$~(1), $40$~(2), $60$~(3), $80$~(4), $100$~(5).
They are shifted by $0.5$ in the vertical direction with respect to the previous one.
}
\label{temi}
\end{figure}

If the parameters of the external potential $\varepsilon_l$ are chosen such that
the values of $\mu_l$ are always within the MI phase,
the density is not perturbed and, therefore, there will be no time dynamics
when the potential is switched off.
In Fig.~\ref{temi}, we show an example of a broader potential,
where local SF regions appear near the center of the potential.
When the potential is switched off, the local inhomogeneities just spread
without creation of any propagating wave packets.
This is consistent with the fact that the sound waves do not exist in the MI phase.

\section{\label{C}Conclusion}

We have studied collective excitations of interacting bosons
in a lattice at zero temperature within the framework of
the {\it gapless} and {\it conserving} time-dependent Gutzwiller approximation.
The excitation modes are calculated within the framework of the linear-response theory
considering small perturbation of the many-body ground state.
We demonstrated that the lowest-energy excitation of the SF has a phonon-like
dispersion relation and derived an analytical expression for the sound velocity
in terms of compressibility and the condensate density which coincides with the hydrodynamic relation.

We have studied the response of the lattice Bose gas in the Bragg scattering process
which provides an experimental tool to observe the excitations.
It is demonstrated that the susceptibility function satisfies the f-sum rule in the whole
parameter region.
Calculations of the transition amplitudes show that within
the Gutzwiller approximation the MI does not respond to the perturbation
caused by the Bragg potential. In the SF phase, we show that only three lowest excitation branches
have significant transition amplitudes, the others being too small to be observed in a real experiment.
The absence of response in the MI phase is a limitation of the Gutzwiller approximation
and can be corrected in the next leading order in the inverse of the coordination number
which would take into account the possibility of a particle-hole pair creation~\cite{Navez,Huber}.

Finally, we have performed simulations of the sound-wave propagation solving numerically
the Gutzwiller equations. The calculations show that sound waves can be created only
in the SF phase and the corresponding velocity is in a good agreement with the
results of the linear-response theory.

\appendix

\section{\label{Der}Derivation of Eq.~(\ref{c_s})}


In order to work out the sound velocity $c_s^0$,
we consider the limit of small $|\vc{k}|$
and look for the lowest energy solution of Eq.~(\ref{evpexc})
as an expansion with respect to $\vc{k}$:
\begin{eqnarray}\label{pert}
\vec{u}_{\bf k}
&=&
\vec{u}_{\bf k}^{(0)}
+
\vec{u}_{\bf k}^{(1)}
+
\vec{u}_{\bf k}^{(2)}+ \dots
\;,
\nonumber\\
\vec{v}_{\bf k}
&=&
\vec{v}_{\bf k}^{(0)}
+
\vec{v}_{\bf k}^{(1)}
+
\vec{v}_{\bf k}^{(2)} +\dots
\;,
\nonumber\\
\omega_{\bf k}
&=&
\omega_{\bf k}^{(0)}
+
\omega_{\bf k}^{(1)}
+
\omega_{\bf k}^{(2)} +\dots
\;.
\end{eqnarray}
The zeroth-order solution satisfies the equation
\begin{equation}
\hbar\omega_{\bf k}^{(0)}
\left(
    \begin{array}{c}
       \vec{u}_{\bf k}^{(0)}\\
       \vec{v}_{\bf k}^{(0)}
    \end{array}
\right)
=
\left(
    \begin{array}{cc}
        A_{0} &  B_{0}\\
       -B_{0} & -A_{0}
    \end{array}
\right)
\left(
    \begin{array}{c}
       \vec{u}_{\bf k}^{(0)}\\
       \vec{v}_{\bf k}^{(0)}
    \end{array}
\right)
\end{equation}
and is non-trivial only in the SF phase with the form:
\begin{equation}
\label{u0}
u_{{\bf k}n}^{(0)}
\equiv
\left(
    n
    +
    \frac{\partial \hbar\omega_0}{\partial \mu}
\right)
c_n^{(0)}
\;,\quad
v_{{\bf k}n}^{(0)}
=
-u_{{\bf k}n}^{(0)}
\;,\quad
\omega_{\bf k}^{(0)}=0
\;.
\end{equation}
The quantities of the first order are governed by the equation
\begin{equation}
\hbar\omega_{\bf k}^{(1)}
\left(
    \begin{array}{c}
       \vec{u}_{\bf k}^{(0)}\\
       \vec{v}_{\bf k}^{(0)}
    \end{array}
\right)
=
\left(
    \begin{array}{cc}
        A_{0} &  B_{0}\\
       -B_{0} & -A_{0}
    \end{array}
\right)
\left(
    \begin{array}{c}
       \vec{u}_{\bf k}^{(1)}\\
       \vec{v}_{\bf k}^{(1)}
    \end{array}
\right)
\;.
\end{equation}
Taking into account the identity
\begin{equation}
\sum_{n'}
\left(
    A_0^{nn'}
    +
    B_0^{nn'}
\right)
\frac{\partial c_{n'}^{(0)}}{\partial\mu}
=
\left(n+\frac{\partial \hbar\omega_0}{\partial \mu}\right)c_n^{(0)}
\;,
\end{equation}
the first-order solution can be written as
\begin{equation}\label{u1}
{u}_{{\bf k}n}^{(1)}
=
{v}_{{\bf k}n}^{(1)}
=
\hbar\omega_\vc{k}^{(1)}
\frac{\partial c_n^{(0)}}{\partial\mu}
\;.
\end{equation}
We substitute all these results in the equation for the quantities of the second order
in $\vc{k}$
\begin{eqnarray}
\label{eqexc2}
&&
\hbar\omega_{\bf k}^{(1)}
\left(
    \begin{array}{c}
       \vec{u}_{\bf k}^{(1)}\\
       \vec{v}_{\bf k}^{(1)}
    \end{array}
\right)
+
\hbar\omega_{\bf k}^{(2)}
\left(
    \begin{array}{c}
       \vec{u}_{\bf k}^{(0)}\\
       \vec{v}_{\bf k}^{(0)}
    \end{array}
\right)
\nonumber\\
&&
=
\left(
    \begin{array}{cc}
        A_{0} &  B_{0}\\
       -B_{0} & -A_{0}
    \end{array}
\right)
\left(
    \begin{array}{c}
       \vec{u}_{\bf k}^{(2)}\\
       \vec{v}_{\bf k}^{(2)}
    \end{array}
\right)
\\
&&
+
\left(
    \begin{array}{cc}
        A_{\bf k}^{(2)} &  B_{\bf k}^{(2)}\\
       -B_{\bf k}^{(2)} & -A_{\bf k}^{(2)}
    \end{array}
\right)
\left(
    \begin{array}{c}
       \vec{u}_{\bf k}^{(0)}\\
       \vec{v}_{\bf k}^{(0)}
    \end{array}
\right)
\;.
\nonumber
\end{eqnarray}
Multiplying Eq.~(\ref{eqexc2}) by the vector
$(\vec{u}_{\bf k}^{(0)}\;,\; -\vec{v}_{\bf k}^{(0)})$
from the left side and taking into account that,
\begin{equation}
\vec{u}_{\bf k}^{(0)} \cdot \vec{u}_{\bf k}^{(0)}
-
\vec{v}_{\bf k}^{(0)} \cdot \vec{v}_{\bf k}^{(0)}
=
0
\;,
\end{equation}
\begin{equation}
\sum_n
\left(
    {u}_{{\bf k}n}^{(0)}
    -
    {v}_{{\bf k}n}^{(0)}
\right)
\frac{\partial c_n^{(0)}}{\partial\mu}
=
\frac{\partial\langle\hat n\rangle}{\partial\mu}
\equiv
\kappa
\;,
\end{equation}
where $\kappa$ is the compressibility, we arrive to Eq.~(\ref{c_s}) for the sound velocity.

\section{\label{Bdr}Bogoliubov's dispersion relation}

We look for the solution of Eq.~(\ref{evpexc}) for arbitrary ${\bf k}$ in the form
\begin{eqnarray}
\label{ab}
\vec{u}_{{\bf k}}
&=&
\vec{u}_{{\bf k}}^{(0)}
a_{\bf k}
+
\vec{u}_{{\bf k}}^{(1)}
b_{\bf k}
\nonumber\\
\vec{v}_{{\bf k}}
&=&
\vec{v}_{{\bf k}}^{(0)}
a_{\bf k}
+
\vec{v}_{{\bf k}}^{(1)}
b_{\bf k}
\end{eqnarray}
where ${u}_{{\bf k}n}^{(0,1)}$ and ${v}_{{\bf k}n}^{(0,1)}$ are given by Eqs.~(\ref{u0}),~(\ref{u1})
with $\omega_{\bf k}^{(1)}$ being replaced by $\omega_{\bf k}$.


Plugging (\ref{ab}) into Eq.~(\ref{evpexc}) and multiplying the resulting 
equations by vectors $\vec{u}_{{\bf k}}^{(0)}$, $\vec{u}_{{\bf k}}^{(1)}$
we obtain linear homogeneous equations for $a_{\bf k}$ and $b_{\bf k}$:
\begin{eqnarray}
\hbar\omega_{\bf k}
\left(
    \vec{u}_{\bf k}^{(0)}
    \cdot
    \vec{u}_{\bf k}^{(1)}
\right)
a_{\bf k}
&=&
\sum_{n,n'}
{u}_{{\bf k}n}^{(1)}
\left(
    A_{\bf k}^{nn'}
    +
    B_{\bf k}^{nn'}
\right)
{u}_{{\bf k}n'}^{(1)}
b_{\bf k}
\nonumber\\
\hbar\omega_{\bf k}
\left(
    \vec{u}_{\bf k}^{(0)}
    \cdot
    \vec{u}_{\bf k}^{(1)}
\right)
b_{\bf k}
&=&
\sum_{n,n'}
{u}_{{\bf k}n}^{(0)}
\left(
    A_{\bf k}^{nn'}
    -
    B_{\bf k}^{nn'}
\right)
{u}_{{\bf k}n'}^{(0)}
a_{\bf k}
\nonumber
\end{eqnarray}
The relations
\begin{equation}
\vec{u}_{\bf k}^{(0)}
\cdot
\vec{u}_{\bf k}^{(1)}
=
\hbar
\omega_{\bf k}
\kappa/2
\;,
\end{equation}
\begin{eqnarray}
&&
\sum_{n,n'}
{u}_{{\bf k}n}^{(1)}
\left(
    A_{\bf k}^{nn'}
    +
    B_{\bf k}^{nn'}
\right)
{u}_{{\bf k}n'}^{(1)}
\\
&=&
\left(
    \hbar\omega_{\bf k}
\right)^2
\left[
    \frac{\kappa}{2}
    +
    \left(
        \frac
        {\partial\psi^{(0)}}
        {\partial\mu}
    \right)^2
    \epsilon_{\bf k}
\right]
\;,
\nonumber
\end{eqnarray}
\begin{equation}
\sum_{n,n'}
{u}_{{\bf k}n}^{(0)}
\left(
    A_{\bf k}^{nn'}
    -
    B_{\bf k}^{nn'}
\right)
{u}_{{\bf k}n'}^{(0)}
=
\epsilon_{\bf k}
{\psi^{(0)}}^2
\;,
\end{equation}
lead us to the result
\begin{eqnarray}
\label{dispersion}
\hbar
\omega_\vc{k}
=
\sqrt
{
 \frac{{2\psi^{(0)}}^2}{\kappa}
 \epsilon_{\bf k}
 +
 \left(\frac{\partial {\psi^{(0)}}^2}{\partial \langle \hat n \rangle}\right)^2
 \epsilon_{\bf k}^2
}
\;.
\end{eqnarray}
In the limit of small $U/J$, $\langle \hat n\rangle ={\psi^{(0)}}^2$
and we recover the well-known Bogoliubov's dispersion relation.
For arbitrary values of $U/J$, Eq.~(\ref{dispersion}) better describes the lowest
excitation branch of the SF than the standard Bogoliubov's dispersion relation
but gives higher values of energies compared to the exact numerical data.

\section{\label{EG}Energy gaps}

We first rewrite Eq.~(\ref{evpexc}) in the form
\begin{equation}
\hbar\omega_{\bf k}
\left(
    \begin{array}{c}
       |{u}_{\bf k}\rangle\\
       |{v}_{\bf k}\rangle
    \end{array}
\right)
=
\left(
    \begin{array}{cc}
        \hat A_{\bf k} &  \hat B_{\bf k}\\
       -\hat B_{\bf k} & -\hat A_{\bf k}
    \end{array}
\right)
\left(
    \begin{array}{c}
       |{u}_{\bf k}\rangle\\
       |{v}_{\bf k}\rangle
    \end{array}
\right)
\;,
\end{equation}
where the operators
\begin{eqnarray}
\hat A_{\bf k}
&=&
-J_{\bf 0}
\psi^{(0)}
(\hat a+{\hat a}^\dagger)
+
\frac{U}{2}
\hat n(\hat n-1)
-\mu \hat n
-\hbar\omega_0
\nonumber\\
&-&
J_{\bf k}
{\hat a}
|s^{(0)}\rangle
\langle s^{(0)}|
{\hat a}^\dagger
+
{\hat a}^\dagger
|s^{(0)}\rangle
\langle s^{(0)}|
\;,
\nonumber\\
\hat B_{\bf k}
&=&
-J_{\bf k}
{\hat a}
|s^{(0)}\rangle
\langle s^{(0)}|
{\hat a}
+
{\hat a}^\dagger
|s^{(0)}\rangle
\langle s^{(0)}|
{\hat a}^\dagger
\;,
\nonumber
\end{eqnarray}
with the ground state $|s^{(0)}\rangle$ defined by Eq.~(\ref{state}),
act on the kets
\begin{eqnarray}
|u_{\bf k}\rangle
=
\sum_n
u_{{\bf k}n}|n\rangle
\;,\quad
|v_{\bf k}\rangle
=
\sum_n
v_{{\bf k}n}|n\rangle
\;.
\end{eqnarray}

The zeroth order exact solution in the small parameter $U/J$ is given by
\begin{eqnarray}
|u_{{\bf k}\lambda}^{(0)}\rangle
&=&
\hat D
\left(
    \psi^{(0)}
\right)
|\lambda\rangle
\;,
\nonumber\\
|v_{{\bf k}\lambda}^{(0)}\rangle
&=&
0
\;,
\end{eqnarray}
where
$\hat D(\alpha)=\exp(\alpha \hat a^\dagger -\alpha^* \hat a)$
is the displacement operator.
This can be shown using the property of the displacement operator
\begin{eqnarray}
\hat D(\alpha)
\hat a
\hat D(\alpha)
=
\hat a + \alpha
\;.
\end{eqnarray}
Finally, first order perturbation theory in small $U/J$ and the relation
\begin{eqnarray}
\langle u_{{\bf k}\lambda}^{(0)}|
\hat B_{\bf k}
|u_{{\bf k}\lambda}^{(0)}\rangle
=0
\end{eqnarray}
valid for $\lambda\geq 2$ allow to establish that
\begin{eqnarray}
\hbar \omega^{(1)}_{{\bf k}\lambda}
=
\langle u_{{\bf k }\lambda}^{(0)}|
\hat A_{\bf k}
|u_{{\bf k }\lambda}^{(0)}\rangle
\;.
\end{eqnarray}
Explicit calculation of the matrix element leads to the result (\ref{gap2})
for the gaps in the excitation spectrum.




\begin{acknowledgments}
This work was supported by the SFB/TR 12 of the German Research Foundation (DFG).
Helpful discussions with Robert Graham are gratefully acknowledged.
\end{acknowledgments}


\end{document}